\documentclass[onecolumn,prd,nofootinbib]{revtex4}
\usepackage{epsf}% Include figure files
\usepackage{dcolumn}% Align table columns on decimal point
\usepackage{bm}% bold math
\usepackage{graphicx}

\usepackage[]{natbib}
\setcitestyle{number,square,comma}
\newcommand{\be}{\begin{equation}}
\newcommand{\ee}{\end{equation}}

\def\4he{$^4$He}
\def\3he{$^3$He}
\def\7li{$^7$Li}

\def\ltsim{\raise 2pt \hbox {$<$} \kern-1.1em \lower 4pt \hbox {$\sim$}}
\def\gtsim{\raise 2pt \hbox {$>$} \kern-1.1em \lower 4pt \hbox {$\sim$}}

\def\sun{\hbox{$_\odot$}}                                    % Sonnensymbol \sun
\def\arcmin{\hbox{$^\prime$}}                               % Bogenminuten
\def\arcsec{\hbox{$^{\prime\prime}$}}                       % Bogensekunden
\def\degr{\hbox{$^\circ$}}                                  % Grad ohne Komma

\def\la{\mathrel{\mathchoice {\vcenter{\offinterlineskip\halign{\hfil
$\displaystyle##$\hfil\cr<\cr\sim\cr}}}
{\vcenter{\offinterlineskip\halign{\hfil$\textstyle##$\hfil\cr
<\cr\sim\cr}}}
{\vcenter{\offinterlineskip\halign{\hfil$\scriptstyle##$\hfil\cr
<\cr\sim\cr}}}
{\vcenter{\offinterlineskip\halign{\hfil$\scriptscriptstyle##$\hfil\cr
<\cr\sim\cr}}}}}                                            % ungefähr kleiner \la
\def\ga{\mathrel{\mathchoice {\vcenter{\offinterlineskip\halign{\hfil
$\displaystyle##$\hfil\cr>\cr\sim\cr}}}
{\vcenter{\offinterlineskip\halign{\hfil$\textstyle##$\hfil\cr
>\cr\sim\cr}}}
{\vcenter{\offinterlineskip\halign{\hfil$\scriptstyle##$\hfil\cr
>\cr\sim\cr}}}
{\vcenter{\offinterlineskip\halign{\hfil$\scriptscriptstyle##$\hfil\cr
>\cr\sim\cr}}}}}                                            % ungefähr größer \ga
\def\zg1{$z\!>\!1$}                                         % z>1 in Fließtext: \zg1
\def\zga1{$z\!\ga\!1$}                                      % z>~1 in Fließtext: \zga1
\def\zsim1{$z\!\sim\!1$}                                    % z~1 in Fließtext: \zsim1
\def\h70{$h_{70}$}                                          % h70 with \h70
\def\hinv70{$h^{-1}_{70}$}                                  % h^-1_70 \hinv70 (h inverse 70)

\begin{document}
\bibliographystyle{abbrvnat}

\title{AGN triggering in the infall regions of distant X-ray luminous galaxy clusters at $\bf 0.9\!<\!z\!\la\!1.6$}

\author{R. Fassbender$^1$, R. \v{S}uhada$^2$,  A. Nastasi$^1$}
\affiliation{$^1$ Max-Planck-Institut f\"ur extraterrestrische Physik (MPE), Postfach 1312, Giessenbachstr., 85741 Garching, Germany \\
%  Max-Planck-Institut fŸr extraterrestrische Physik, Postfach 1312, Giessenbachstr., 85741 Garching, Germany
              $^2$ Department of Physics, Ludwigs-Maximilians Universit\"at M\"unchen, Scheinerstr. 1, 81679  Munich, Germany} 
\affiliation{}

\date{{\today}\\}

\begin{abstract}
\noindent
Observational constraints on the average radial distribution profile of AGN in distant galaxy clusters can provide important clues on the triggering mechanisms of AGN activity in dense environments and are essential for a completeness evaluation of cluster selection techniques in the X-ray and mm-wavebands. The aim of this work is  a statistical study with XMM-{\it Newton} of the presence and distribution of X-ray AGN in the large-scale structure environments of 22 X-ray luminous galaxy clusters in the redshift range $0.9\!<\!z\!\la\!1.6$ compiled by the XMM-{\it Newton} Distant Cluster Project (XDCP). 
To this end, the X-ray point source lists from detections in the soft-band (0.35-2.4\,keV) and full-band (0.3-7.5\,keV) were stacked in cluster-centric coordinates and compared to average background number counts extracted from three independent control fields in the same observations. A significant full-band (soft-band) excess of $\sim$78 (67) X-ray point sources is found  in the cluster fields within an angular distance of 8\arcmin \ (4\,Mpc)  at a statistical confidence level of 4.0\,$\sigma$ (4.2\,$\sigma$), % in the soft-band (full-band), 
corresponding to an average number of detected excess AGN per cluster environment of 3.5$\pm$0.9 (3.0$\pm$0.7). The data point towards a rising radial profile in the cluster region (r$<$1\,Mpc) of predominantly low-luminosity AGN with an average detected excess of about one point source per system,
 %with a tentative preferred 
 with a tentative preferred occurrence %tentatively preferentially located 
 along the main cluster elongation axis. A second statistically significant overdensity of brighter soft-band detected AGN is found at cluster-centric distances of 4\arcmin-6\arcmin (2-3\,Mpc), corresponding to about three times the average cluster radius R$_{200}$ of the systems. If confirmed, these results would support the idea of two different physical triggering mechanisms of X-ray AGN activity in dependence of the radially changing large-scale structure environment of the distant clusters. 
%Concerning
For high-$z$ cluster studies at lower spatial resolution with the upcoming eROSITA all-sky X-ray survey, the results suggest that cluster-associated X-ray AGN may impose a bias in the spectral analysis of high-$z$ systems, while their detection and flux measurements in the soft-band may not be significantly affected.

\end{abstract}

\pacs{}

\keywords{}

\maketitle

%%%%%%%%%%%%%%%%%%%%%%%%%%%%%%%%%%%%%%%%%%%%%%%%%%%%%%%%%%%%%%%%%%%%%%%%%%%
%%%%%%%%%%%%%%%%%%%%%%%%%%%%%%%%%%%%%%%%%%%%%%%%%%%%%%%%%%%%%%%%%%%%%%%%%%%

\section{~Introduction}

\noindent
Observational studies of the connection of Active Galactic Nuclei (AGN) with the large-scale environment of massive clusters of galaxies  
and their mutual cosmic evolution can provide important insights into the physical conditions necessary to trigger or suppress AGN activity in galaxies. In this respect, AGN activity can be charted in dependence of  the changing environments of  galaxy clusters as a function of cluster-centric distance: from the dense cores, to the cluster outskirts, and further out to the matter infall regions and the surrounding cosmic web.  Furthermore, the evolution of the occurrence of X-ray and radio AGN in cluster environments as a function of redshift is of key importance for the characterization and completeness evaluation of ongoing and future high-$z$ cluster surveys  in the X-ray band and via the Sunyaev-Zeldovich effect (SZE) at mm wavelengths.

Numerous  {\it Chandra}  sample studies on the  X-ray AGN content of galaxy clusters up to redshifts of about unity \cite[e.g.][]{Tomczak2011a,Martini2009a,Gilmour2009a,Branchesi2007a,Eastman2007a,Cappelluti2005a,Ruderman2005a} have firmly established that the AGN fraction in clusters is significantly rising as a function of redshift. On the other hand, the cluster environment appears to suppress the occurrence of X-ray AGN activity in massive galaxies compared to a field  galaxy sample at all probed redshifts so far \cite[e.g.][]{Koulouridis2010a,Martini2009a,Eastman2007a} and the distribution of X-ray AGN is significantly less concentrated  in terms of cluster-centric distance in comparison to radio AGN  \cite[e.g.][]{Rumbaugh2011a,Branchesi2007a,Johnson2003a}. However, owing to the small number of very distant test clusters at $z\!>\!0.9$ available to these studies, the  data basis  at redshifts beyond unity is still very sparse, and at $z\!>\!1.3$ X-ray cluster studies are limited to a single systems \cite{Hilton2010a,Pierre2011a}.
So far, the only cluster AGN study with a sizable sample of systems at the epoch $1\!<\!z\!\la\!1.5$ was presented in \cite{Galametz2009a} based on infrared selected clusters from the IRAC Shallow Cluster Survey \cite{Eisenhardt2008a}, which found a continuing trend of increasing AGN fractions with redshifts in systems with an average halo mass of $\sim$$10^{14}\,\mathrm{M_{\sun}}$.

%\citep{Koulouridis2010a}
%AGN as tracers of the large-scale structure (LSS)
%low-z results

%Individual Clusters:
%\cite{Johnson2003a}
%\cite{Hilton2010a}
%
%LSS:
%\cite{Rumbaugh2011a}
%
%Mid-z:
%\cite{Cappelluti2005a}
%\cite{Martini2009a}
%\cite{Branchesi2007a}
%
%z=0.3-0.7
%\cite{Eastman2007a}
%\cite{Ruderman2005a}
%

%Most Extensive Study
%\cite{Gilmour2009a}

%high-z, MIR selected:
%\cite{Galametz2009a}
%\citep{Tomczak2011a}

%XLF:
%\cite{Yang2009a}

%Radio AGN in Groups
%\citep{Smolcic2011a}

The aim of this paper is to extend  the accessible redshift regime for a sample of X-ray selected clusters to  $0.9\!<\!z\!\la\!1.6$ in order to perform a  statistical study of the X-ray point source excess and its radial dependence in distant  X-ray luminous systems.
In contrast to the previous studies based on targeted follow-up observations with  {\it Chandra}, this work is %largely 
built upon archival   
XMM-{\it Newton} observations, in which the clusters have been serendipitously detected. This distant cluster sample and the performed X-ray point source stacking analysis is introduced in Sect.\,\ref{s2_Sample}.  The results are presented in Sect.\,\ref{s4_Results},  followed by the discussion in Sect.\,\ref{s5_Discussion}, and conclusions in  Sect.\,\ref{s6_Summary}.
A standard $\Lambda$CDM cosmological model with parameters ($H_0$, $\Omega_{\mathrm{m}}$, $\Omega_{\mathrm{DE}}$, w)=(70\,km\,s$^{-1}$Mpc$^{-1}$, 0.3, 0.7, -1) is assumed throughout this paper, which yields a median projected angular scale of 8.2\,kpc/\arcsec \ (corresponding to 2.0\arcmin/Mpc) with $<$5\% variation in the probed redshift interval $z\!=\!0.9$-1.6.

%0.9: gives a scale of 7.791 kpc/"
%1.1: gives a scale of 8.172 kpc/"
%1.2: gives a scale of 8.293 kpc/"
%1.6: This gives a scale of 8.471 kpc/"

%aim of this work:
%statistical analysis of the environment of X-ray clusters with the attempt to identify overdensity of cluster associated X-ray AGN 

%want to answer the question on whether there is an excess of X-ray point sources around distant cluster, and if yes

%cosmological angular scale factor in range 0.9-1.6

%%%%%%%%%%%%%%%%%%%%%%%%%%%%%%%%%%%%%%%%%%%%%%%%%%%%%%%%
\section{~Distant Cluster Sample and X-ray Stacking Analysis}
\label{s2_Sample}
\label{s3_Analysis}

%------------------------------------------------------------------------------------------------------------------------------------------
\subsection{~Cluster Sample}
\label{s2_SampleDescription}
\noindent
This work uses the largest published sample of distant X-ray luminous galaxy clusters at redshifts $z\!>\!0.9$ to date as presented in 
\citet{Fassbender2011c}. The sample comprises 22 X-ray selected clusters in the redshift range $0.9\!<\!z\!\la\!1.6$ with a median system mass of M$_{200}\!\simeq\!2 \times 10^{14}$\,M$_{\sun}$. All clusters are part of the XMM-{\it Newton} Distant Cluster Project (XDCP, e.g. \cite{Fassbender2011c}, \cite{RF2007PhD}, \cite{HxB2005a}), a serendipitous X-ray survey focussing on the detection and study of galaxy clusters in the first half of cosmic time. Table\,\ref{tab_Cluster_Sample} provides an overview of the considered cluster sample, including the  XMM-{\it Newton} observations used for this study, their effective clean exposure time\footnote{We define the ECT as the period during which all three instruments in imaging operation would collect the equivalent number of soft science photons for the particular observation.} (ECT) after flare removal, and original relevant source publications for the individual clusters.

%XDCP
%Current sample
% Why this sample is a good starting point

All clusters in this sample have by construction medium to deep XMM-{\it Newton}  observations available, whereas the  {\it Chandra} archive coverage of this newly constructed sample is currently less than 30\%. The complete coverage and high sensitivity is the main advantage to perform a first X-ray point source study around distant X-ray clusters with XMM-{\it Newton}. A second advantage is the
%which also offers a
 larger 30\arcmin-diameter field-of-view (FoV), which allows direct measurements of background counts in the same observation.   
XMM-{\it Newton}'s spatial resolution of 5-15\arcsec \ (FWHM), on the other hand, is significantly lower than for {\it Chandra}, which implies that the central cluster core regions of up to $\sim$15\arcsec$\simeq$120\,kpc cannot be properly probed for point sources in addition to the underlying detected extended X-ray emission of the cluster.

%requirements:
%full X-ray coverage of cluster environments
%high sensitivity

In any case, the statistical detection of a point source excess around distant clusters at $z\!>\!0.9$ is a challenging task since the average background density of X-ray point sources in the considered  XMM-{\it Newton} fields is more than 20 within a 6\arcmin-radius ($\sim$3\,Mpc) compared to an expected cluster excess of a few sources ($\sim$1.5 for $z\!<\!0.9$ systems \cite{Gilmour2009a}).
On the other hand, any measurable point source excess associated with $z\!>\!0.9$ clusters can be directly attributed to AGN activity with X-ray luminosities of L$_{\mathrm{X}}\!>\!10^{43}$\,erg\,s$^{-1}$ owing to the average soft-band point source detection limit in  
the XMM-{\it Newton} fields of $\ga$$10^{-15}$\,erg\,s$^{-1}$cm$^{-2}$.

%advantages
%AGN density should be higher
%
%possibilities and challenges at high-z

% ------------------------------- TABLE with Cluster Sample --------------------------------------------------------------
% 
%:MASTER_TABLE
\begin{table*}[t]
%\fulltable
\caption{List of the 22 distant galaxy clusters at  $z\!>\!0.9$ from \citet{Fassbender2011c} used for the stacking analysis in this work.
The table lists a cluster identification number (column 1), the system redshift (2),  the official cluster name (3), X-ray centroid coordinates (4+5), the observation identification number of the XMM-{\it Newton} field used  (6), the corresponding effective clean time (ECT) of the field in (7),  and relevant literature references  to the cluster in (8). 
}\label{tab_Cluster_Sample}
\centering
\begin{tabular}{ l l l l l l l l}
\hline \hline     
ID & $z$ & Official Name &  RA & DEC &  OBSID & ECT &  References \\
 &  &   & J2000 & J2000  &  & ksec &     \\
(1)  &  (2) &  (3)  & (4)  & (5)  & (6)  & (7)  & (8)     \\

\hline

C01 &	1.579 &		XDCP\,J0044.0-2033 &		00:44:05.2 &	-20:33:59.7 & 0042340201 & 8.5  & \cite{Santos2011a}  \\
C02 &	1.555 &		XDCP\,J1007.3+1237 &	 	10:07:21.6 & 	+12:37:54.3 & 0140550601 & 19.4 & \cite{Fassbender2011a} \\
C03 &	1.490 &		XDCP\,J0338.8+0021 &		03:38:49.5 &	+00:21:08.1& 0036540101 & 18.0  &  \cite{Nastasi2011a} \\
C04 &	1.457 &		XMMXCS\,J2215.9-1738 &	22:15:58.5 &	-17:38:05.8  & 0106660101 & 51.7 &  \cite{Stanford2006a}, \cite{Hilton2007a,Hilton2009a,Hilton2010a}\\
C05 &	1.396 &		XDCP\,J2235.3-2557 &		22:35:20.4 &	-25:57:43.2 & 0111790101& 13.6 &   \cite{Mullis2005a},\cite{Rosati2009a} \\
C06 &	1.358 &		XDCP\,J1532.2-0837 &		15:32:13.2 &	-08:37:01.4  & 0100240801 & 22.4 &  \cite{Suhada2011a} \\
C07 &	1.335 &		SpARCS\,J0035.8-4312 &	00:35:50.1 &	-43:12:10.3  & 0148960101 & 47.2 & \cite{Wilson2009a}, \cite{Fassbender2011c} \\
C08 &	1.237 &		RDCS\,J1252.9-2927 &		12:52:54.5 &	-29:27:18.0  & 0057740401 & 62.0&  \cite{Rosati2004a}\\ 
C09 &	1.227 &		XDCP\,J2215.9-1751 &		22:15:56.9 &	-17:51:40.9  & 0106660601 & 82.2 &  \cite{DeHoon2011a}\\
C10 &	1.185 &		XDCP\,J0302.1-0001 &		03:02:11.9 &	-00:01:34.3  & 0041170101 & 40.9 &  \cite{Suhada2011a} \\
C11 &	1.122 &		XDCP\,J2217.3+1417 &		22:17:20.8 &	+14:17:54.6 & 0103660301 & 10.3 &  \cite{Fassbender2011c} \\
C12 &	1.117 &		XDCP\,J2205.8-0159 &		22:05:50.3 &	-01:59:27.4  &0012440301 & 24.9 & \cite{Dawson2009a}, \cite{Fassbender2011c} \\
C13 &	1.097 &		XDCP\,J0338.7+0030 &		03:38:44.2 &	+00:30:01.8  &  0036540101 & 18.0 & \cite{Pierini2011a}\\
C14 &	1.082 &		XDCP\,J1007.8+1258 &		10:07:50.5 &	+12:58:18.1 & 0140550601 & 19.4 & \cite{Schwope2010a} \\
C15 &	1.053 &		XLSS\,J0227.1-0418 &		02:27:09.2 &	-04:18:00.9  & 0112680101& 22.7 &  \cite{Pacaud2007a}, \cite{Adami2011a} \\
C16 &	1.050 &		XLSS\,J0224.0-0413 &		02:24:04.1 &	-04:13:31.7 & 0112680301 & 19.2 &  \cite{Pacaud2007a}, \cite{Maughan2008a}, \cite{Adami2011a}\\
C17 &	1.000 &		XDCP\,J2215.9-1740 &		22:15:57.5 &	-17:40:25.6 & 0106660101 & 51.7 &  \cite{DeHoon2011a} \\
C18 &	0.975 &		XDCP\,J1229.4+0151 &		12:29:29.2 &	+01:51:31.6 & 0126700201 & 8.7 & \cite{Santos2009a} \\
C19 &	0.975 &		XDCP\,J1230.2+1339 &		12:30:16.9 &	+13:39:04.3 & 0112552101 & 10.3 &  \cite{Fassbender2011b} \\
C20 &  	0.959 &   	        XDCP\,J0027.2+1714 &         	00:27:14.3 &  	+17:14:36.3 & 0050140201 & 41.8 &  \cite{Fassbender2011c} \\
C21 &	0.947 &		XDCP\,J0104.3-0630 &		01:04:22.3 &	-06:30:03.1 & 0112650401 & 18.4 & \cite{RF2008b}  \\
C22 &	0.916 &		XDCP\,J0338.5+0029 &		03:38:30.5 &	+00:29:20.2 &  0036540101 & 18.0 &  \cite{Fassbender2011c}  \\

\hline
\end{tabular}
\end{table*}

%\endfulltable
%--------------------------------------------------------------------------------------------------------------------------

%%%%%%%%%%%%%%%%%%%%%%%%%%%%%%%%%%%%%%%%%%%%%%%%%%%%%%%%
%\section{~X-ray Stacking Analysis}
%\label{s3_Analysis}

%------------------------------------------------------------------------------------------------------------------------------------------
\subsection{~X-ray Data}
\label{s2_XrayData}

\begin{figure}[t]
\centering
\includegraphics[angle=0,clip,width=0.495\textwidth]{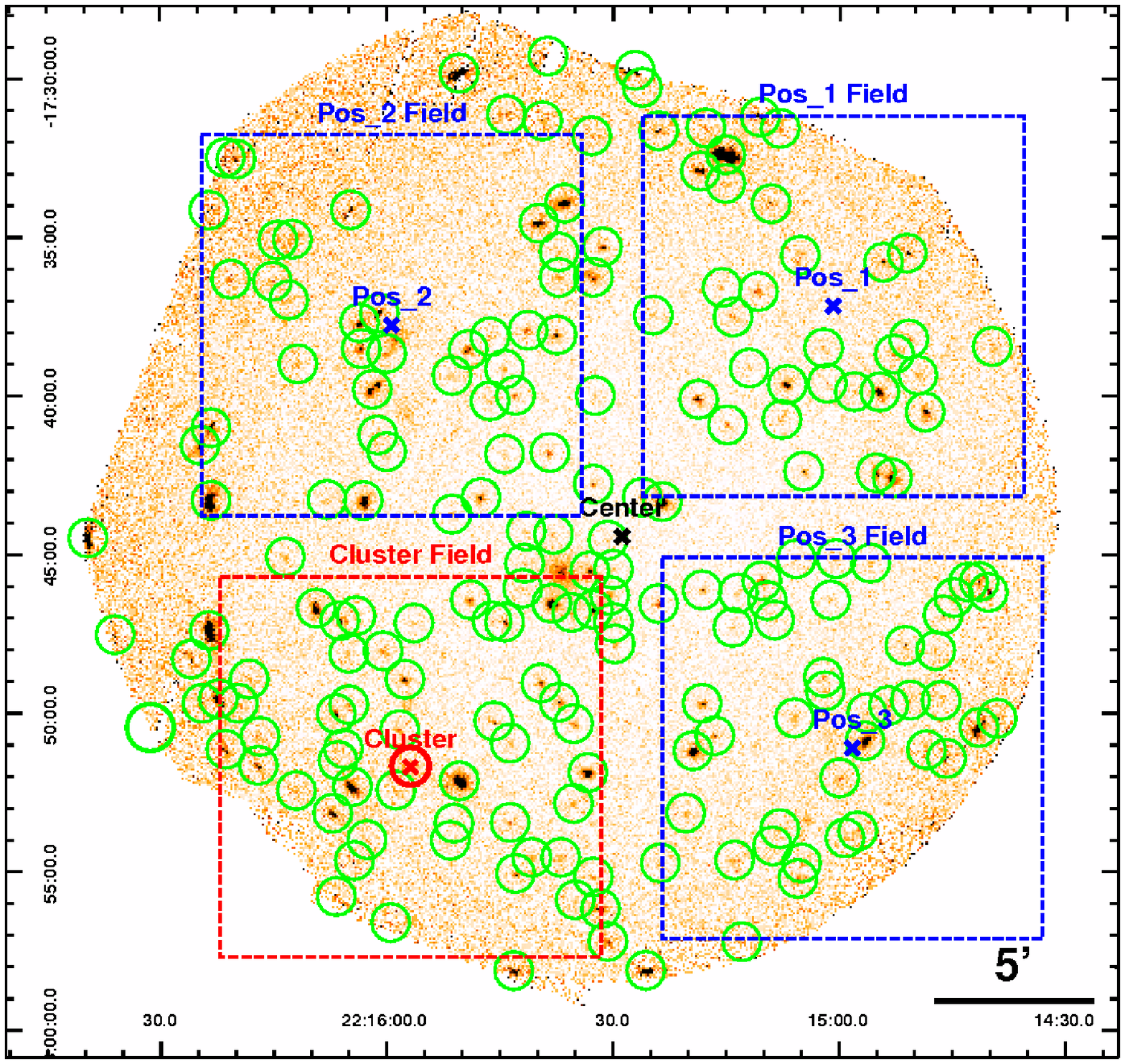}
\includegraphics[angle=0,clip,width=0.495\textwidth]{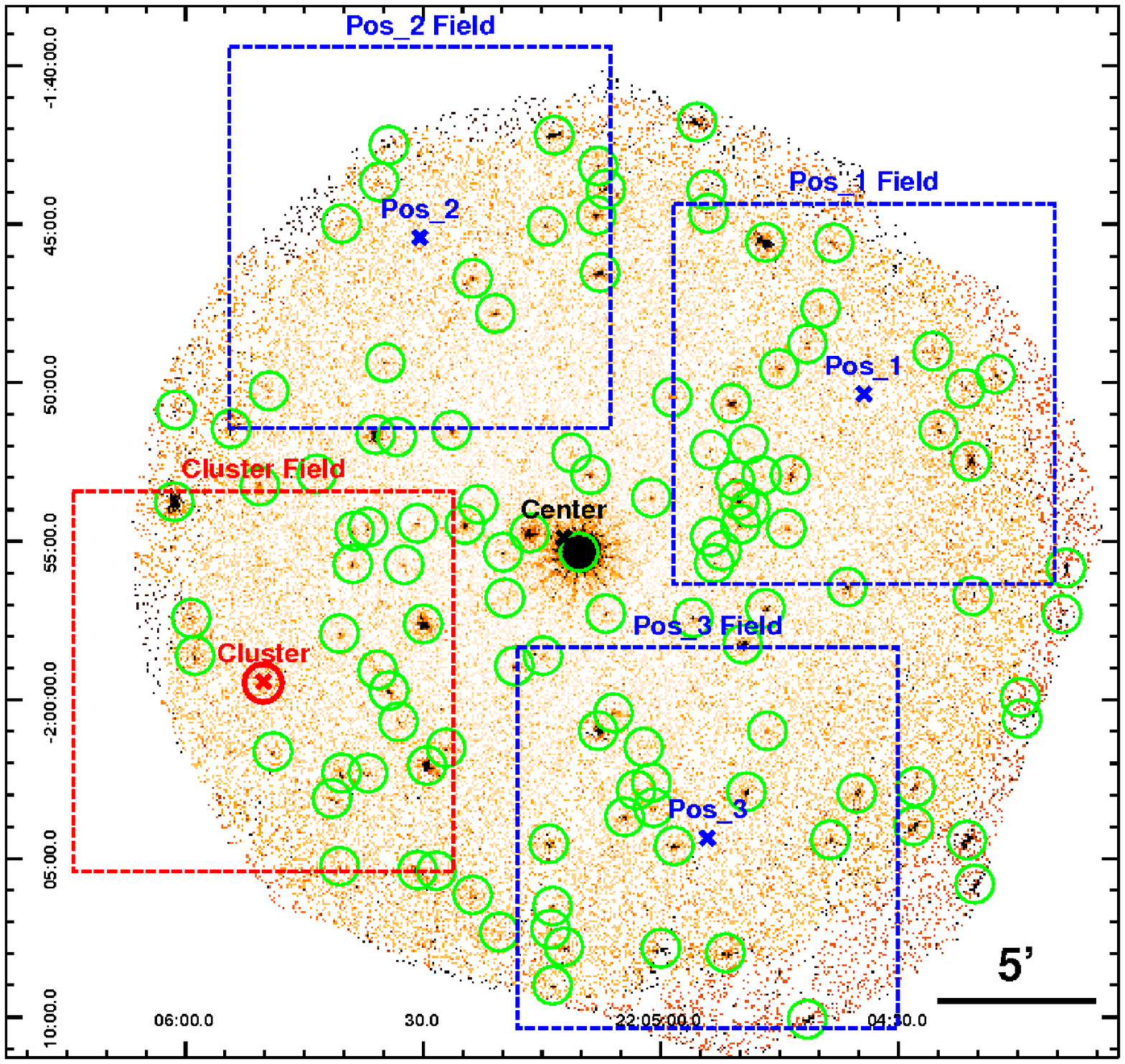}
%\includegraphics[angle=0,clip,width=0.49\textwidth]{Pub_v2_FoV_Extraction_X2215b.ps}
%\includegraphics[angle=0,clip,width=0.49\textwidth]{Pub_FoV_Extraction_X2205.ps}
%\vspace{-1ex}
\caption{Examples of the extraction and stacking process for the 82.2\,ksec field of cluster XDCP\,J2215.9-1751 (C09)  in the left panel and the 24.9\,ksec observation of the system XDCP\,J2205.8-0159 (C12) in the right panel. The cluster positions and the 12\arcmin$\times$12\arcmin \ region around them are indicated in red. The three control fields at equivalent off-axis angles from the center position (black) are marked in blue. Green circles mark the positions of all detected soft-band X-ray point sources in the fields.}
\label{f1_Extraction_FoV}       % Give a unique label
\end{figure}

\noindent
The XMM-{\it Newton} observations used for this work are in general the original discovery fields of the clusters as listed in 
Table\,\ref{tab_Cluster_Sample} (column 6), with the exception of C08 for which the deeper follow-up observation was used. 
The clean effective exposure time of these fields after flare removal (column 7) ranges from 8.5-82\,ksec with a median integration time of  19\,ksec. Three of the fields contain 2-3 confirmed distant clusters inside their FoV, which are treated here as independent observations for each cluster source. 

Most of the identified high-$z$ clusters are located at off-axis angles  $\Theta_{\mathrm{off}}$ between 5-12\,arcmin from the optical axis, which was defined as the aimpoint of the  PN  (the most sensitive instrument)  labelled as center in Fig.\,\ref{f1_Extraction_FoV}.
The 30\arcmin-diameter FoV of XMM-{\it Newton} allows to define three quasi independent background control regions inside the same XMM-{\it Newton} observation at the same off-axis angle with  centers rotated by 90, 180, and 270 degrees about the optical axis as is shown in   Fig.\,\ref{f1_Extraction_FoV}. 
{These fixed rotation angles to the control field centers ensure maximally spaced distances at a given off-axis angle between control and clusters fields in order to maximize the  non-overlapping regions between them.}
When comparing the cluster environment and background field point source counts, this approach {(to first order)} automatically accounts for (i) changes in the effective exposure time across the FoV due to vignetting, (ii) the changing PSF as a function of off-axis angle, and (iii) possible incomplete coverage beyond some cluster-centric distance in the outer radial direction due to the edge of the instrumental FoV.  For the observations in  which the cluster location is at small off-axis angles of  $\Theta_{\mathrm{off}}\!<\!8\arcmin$ the control field locations were shifted back to  $\Theta_{\mathrm{off}}\!=\!8\arcmin$ at their respective rotation angles of 90, 180, and 270 degrees. This ensures that cluster and background fields are not overlapping within their minimum  separations of $>\!8\arcmin$, while only slightly changing the average effective sensitivity in the control fields relative to the cluster environment.

The X-ray source detection was performed as part of the XDCP distant cluster survey as detailed in  \citet{Fassbender2011c}. Here the focus of the analysis lies on the detected point sources in the individual cluster fields, which are displayed by green circles in Figs.\,\ref{f1_Extraction_FoV}\,\&\,\ref{f2_Extraction_Cluster}. The X-ray source lists were produced with the XMM Science Analysis Software\footnote{\url{http://xmm.esac.esa.int/sas/}} (SAS) tasks  {\tt eboxdetect} for a first sliding box detection of candidate sources followed by a maximum likelihood fitting procedure with  {\tt emldetect} for the final source parameter determination.
For this work, the source lists of two different detection procedures are used: (1) a  {\it soft-band} source detection in the energy band 0.35-2.4\,keV down to point source significances of about 3\,$\sigma$ and (2) a {\it full-band} detection in multiple bands covering the energy range 0.3-7.5\,keV down to point source significances of about 2\,$\sigma$. Although originally developed and optimized for distant cluster detections, the two schemes are able to distinguish and probe different aspects of the AGN-cluster connection. The 
 investigation in the {\it soft-band}  is mostly sensitive to low-absorption type-I AGN and can probe the effects of cluster-AGN on the detection efficiency of distant clusters at lower spatial resolution with the upcoming all-sky survey  eROSITA \cite{Predehl2010a} as discussed in Sect.\,\ref{s5_eROSITA_implic}.
The  {\it full-band} detection, on the other hand, is sensitive to the full type-I  and type-II AGN population and can hence provide a more complete census of AGN activity in the vicinity of distant clusters. 
 
In order to allow a robust determination of the pure point source excess, all 105 detected {\it extended} X-ray sources in the considered fields were removed from the master source lists, including all of the 22 distant cluster targets listed in  Table\,\ref{tab_Cluster_Sample}.
Additionally, the {\it soft-} and  {\it full-band}  master lists were manually cleaned from obvious spurious detections ($\simeq$5.5\%) in the immediate vicinity of very bright X-ray sources in the FoV, such as the central source displayed in the right panel of Fig.\,\ref{f1_Extraction_FoV}. The final combined clean X-ray point source lists contain 2770 objects for the {\it soft-band} and 4228 sources for the {\it full-band} detection. With an effective total X-ray coverage of 4.246\,deg$^2$ for the considered fields, the average point source surface densities amount to 0.182\,srcs/arcmin$^2$ in the  
 {\it soft-band} and 0.277\,srcs/arcmin$^2$ in the  {\it full-band}.

%------------------------------------------------------------------------------------------------------------------------------------------
\subsection{~Point Source Stacking Analysis}
\label{s2_StackinAna}

\noindent
In order to keep the  analysis simple, the  focus is placed on a possible {\em detectable} AGN point source excess in cluster environments with respect to control fields, irrespective of the flux and luminosity distribution of such sources. Owing to the varying effective exposure time of the different  XMM-{\it Newton} fields in Table\,\ref{tab_Cluster_Sample} (columns 6+7), the point source detection sensitivities scale with roughly the inverse square root of the exposure time and vary across the FoV as a function of off-axis angle due to increased vignetting effects at larger $\Theta_{\mathrm{off}}$. The typical 0.5-2.0\,keV soft-band point source sensitivities of  $\simeq\!(1$-$2)\!\times\!10^{-15}$\,erg\,s$^{-1}$cm$^{-2}$ ensure, however, that any detected point source excess 
associated with clusters in the redshift regime at $0.9\!<\!z\!\la\!1.6$ can be safely attributed to X-ray AGN\footnote{Under the well-justified assumption that a statistically significant, background-subtracted population of  excess counts is physically associated with the LSS cluster environments at the respective redshifts.} , rather than e.g.~star-forming galaxies at lower X-ray luminosities. 
%This conclusion is 

For the point source stacking analysis the following approach is adapted: (a) For each of the 22 distant clusters in the considered sample a sub-image is extracted  from the corresponding XMM-{\it Newton} observation that places the cluster at the central image coordinates (X$_{\mathrm{cen,cl}}$ , Y$_{\mathrm{cen,cl}}$) as displayed in red in Fig.\,\ref{f1_Extraction_FoV}. The same procedure is repeated at the three control field positions at rotation angles of 90, 180, and 270 degrees about the optical axis shown in blue in the same figure. (b) In order to ensure a homogeneous effective exposure time distribution in the final stacked cluster and control fields, the identical sub-images at the four positions as in the first step are extracted from the associated exposure maps. (c)  The X-ray source lists for the   {\it soft-band}  and {\it full-band}  detections are loaded to each of the four associated extracted sub-frames (one cluster plus three control positions) in the world-coordinate system (WCS) and re-saved in the image coordinate system of each frame.
(d) For each of the four positions per cluster, the 22 extracted sub-images and exposure maps around the cluster position and control fields are co-added to result in the final deep image and exposure map stacks for each position. (e) Similarly, the 22 source lists in image coordinates at each position are concatenated into single master source files at each cluster and control field position for the 
{\it soft-band}  and {\it full-band}  detections separately. The source distribution of these stacked master files can now be further analyzed as a function of distance to the central extraction position with image coordinates  (X$_{\mathrm{cen,cl}}$ , Y$_{\mathrm{cen,cl}}$).

\begin{figure}[t]
\centering
\includegraphics[angle=0,clip,width=0.33\textwidth]{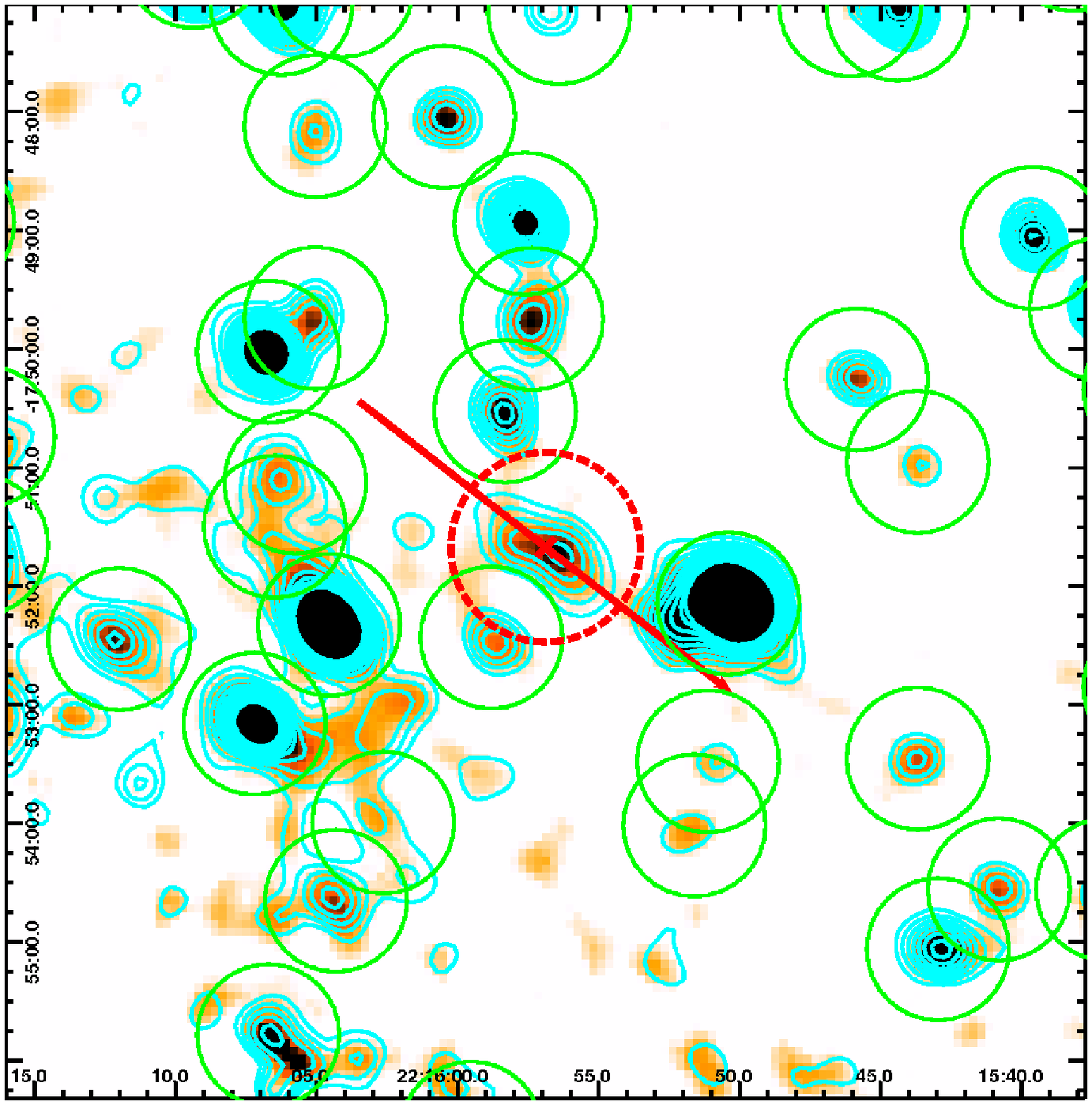}
\includegraphics[angle=0,clip,width=0.33\textwidth]{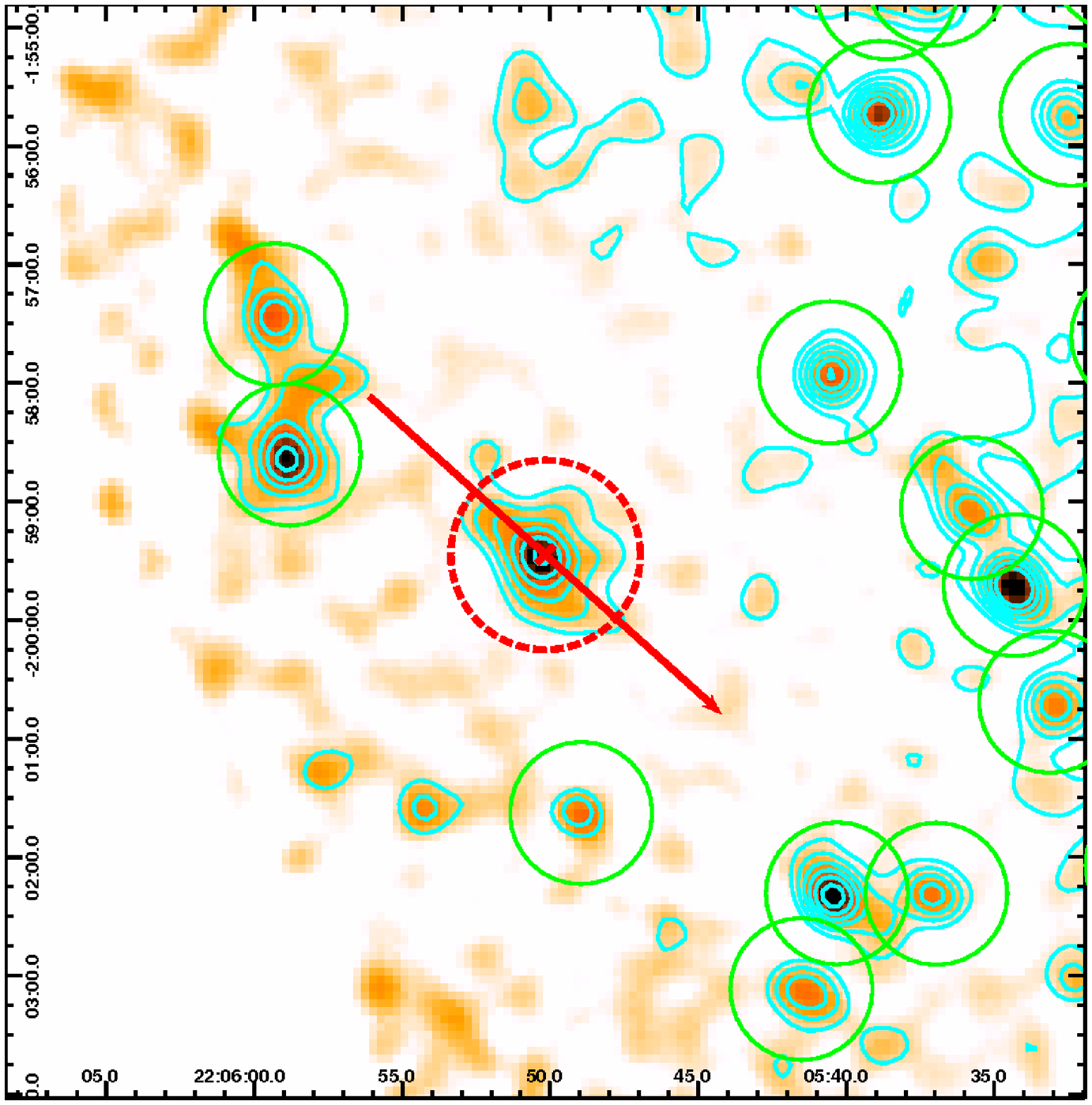}
\includegraphics[angle=0,clip,width=0.33\textwidth]{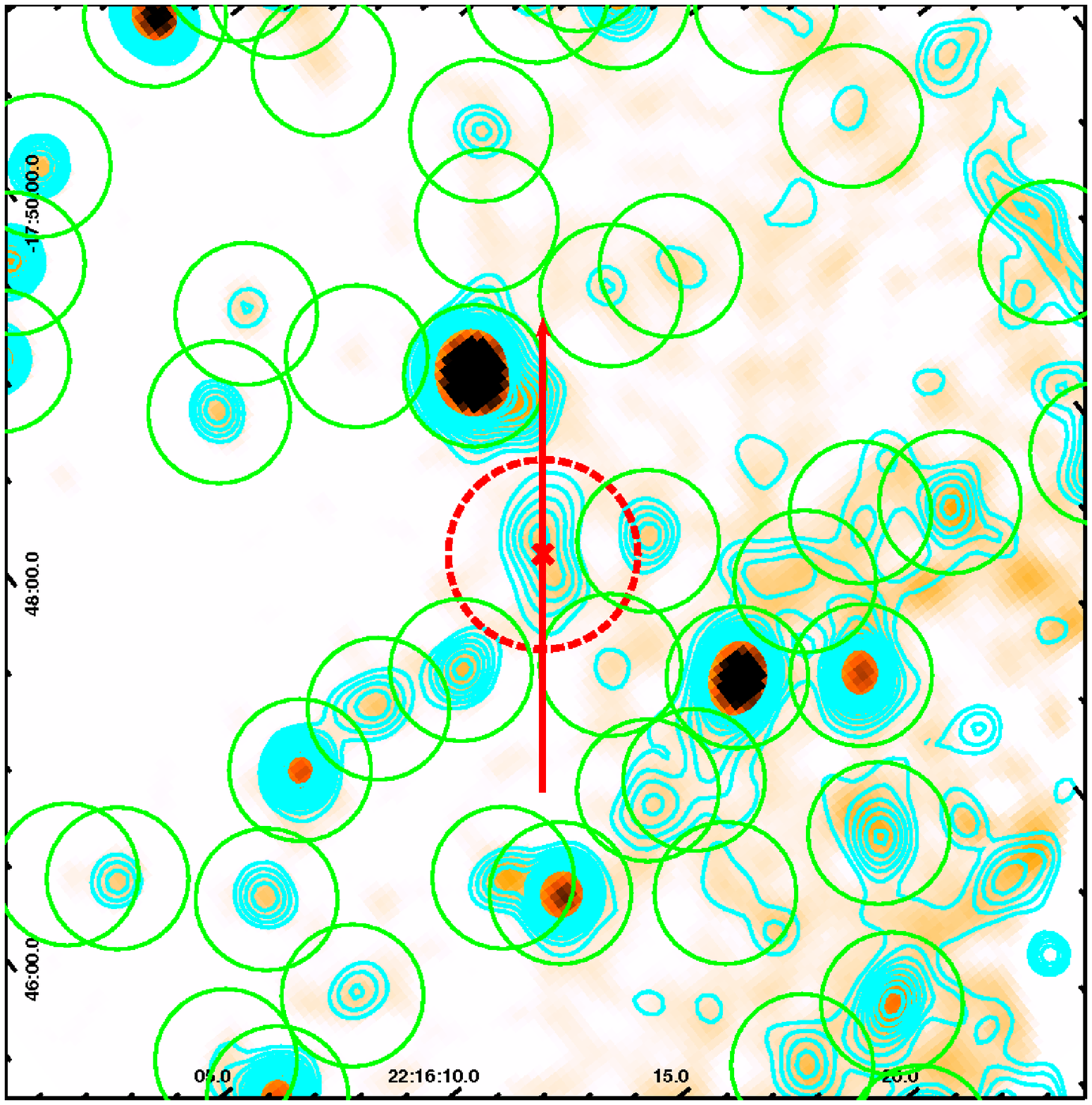}
\includegraphics[angle=0,clip,width=0.33\textwidth]{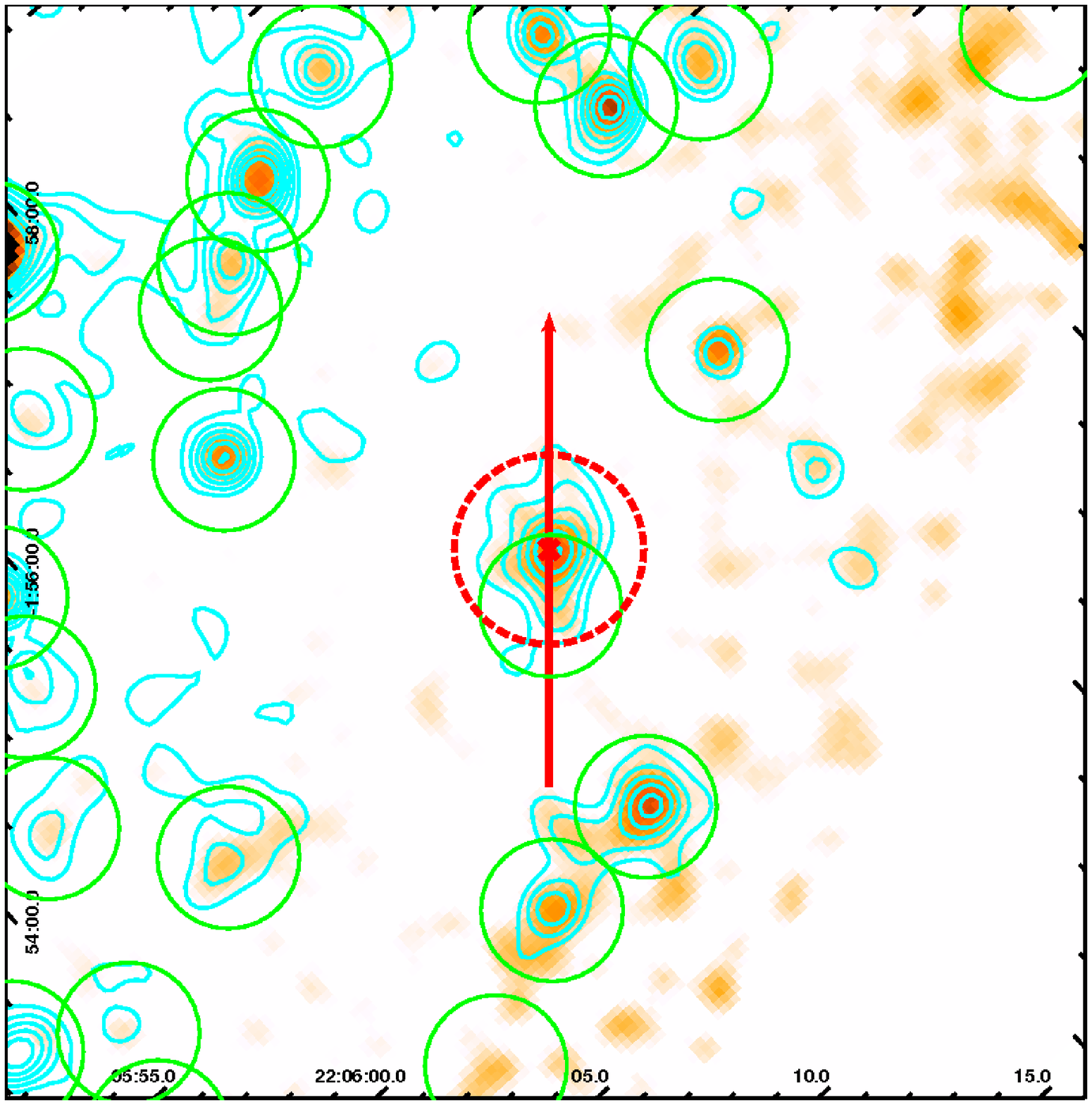}

\vspace{-1ex}
\caption{Close-up examples of the stacking process for the 9.2\arcmin$\times$9.2\arcmin cluster environments of  XDCP\,J2215.9-1751 (C09)  in the left panels and XDCP\,J2205.8-0159 (C12) in the right panels. Cluster positions are marked by red crosses and dashed circles, and the main elongation axis of the ICM emission is indicated by the red arrow with 2\arcmin \ length in both directions from the cluster centers. The top panels show the smoothed soft-band image and the 0.35-2.4\,keV detected sources (green circles) in their original orientation (North up, East to the left). The bottom panels display the rotated images and detected sources in the full 0.3-7.5\,keV energy range with rotation angles that align the cluster elongation axis in the North-South direction. All panels show overlaid  the soft-band X-ray surface brightness contours in cyan.}
\label{f2_Extraction_Cluster}       % Give a unique label
\end{figure}

The extraction procedure with the cluster (or control field) position in the center of the sub-image is depicted in the top panels of  Fig.\,\ref{f2_Extraction_Cluster} for the two systems C09 (left) and C12 (right). The detected extended X-ray emission of the cluster (red dashed circles) does by definition not contribute to the point source statistics (green circles). In order to also investigate potentially preferred directions of the AGN excess in cluster environments, a second variant of the previously described stacking analysis is performed. This time the visually determined main elongation axis of the detected cluster emission (red arrows in Fig.\,\ref{f2_Extraction_Cluster}) is assumed to be a proxy for the main matter infall and assembly axis of the cluster   \cite[e.g.~see][]{Fassbender2011b}. All extracted sub-images and source lists of steps (a-c) are then rotated about the central position with coordinates  (X$_{\mathrm{cen,cl}}$ , Y$_{\mathrm{cen,cl}}$) in a way to align the cluster elongation axis for each of the 22 cluster fields in the North-South direction as shown in the lower panels of 
Fig.\,\ref{f2_Extraction_Cluster}. Subsequently, step (e) is repeated with these rotated source lists to produce a stacked master catalog of X-ray sources in a coordinate system where the cluster elongation axes are co-aligned in the vertical direction.     

%Stacking strategy
%Extraction of cluster and control fields
%Image and source list stacking

%Rotations along main elongation axis of clusters

\begin{figure}[t]
\centering
\includegraphics[angle=0,clip,width=0.49\textwidth]{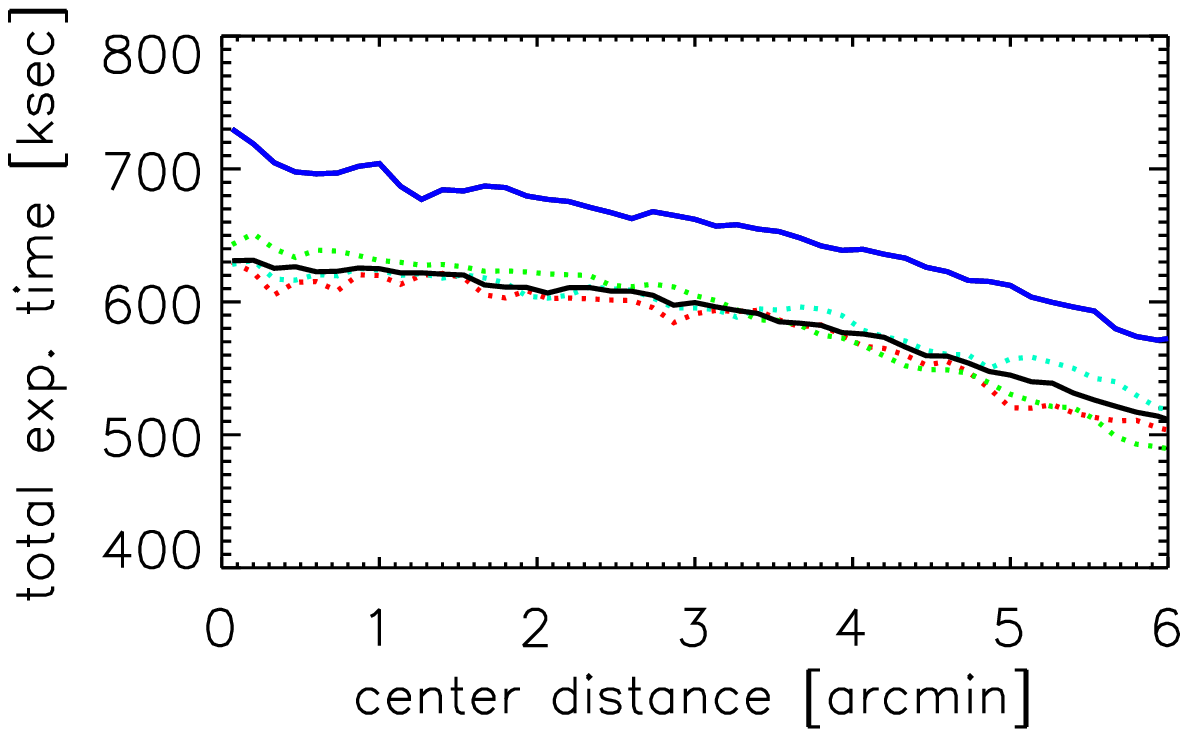}
\includegraphics[angle=0,clip,width=0.49\textwidth]{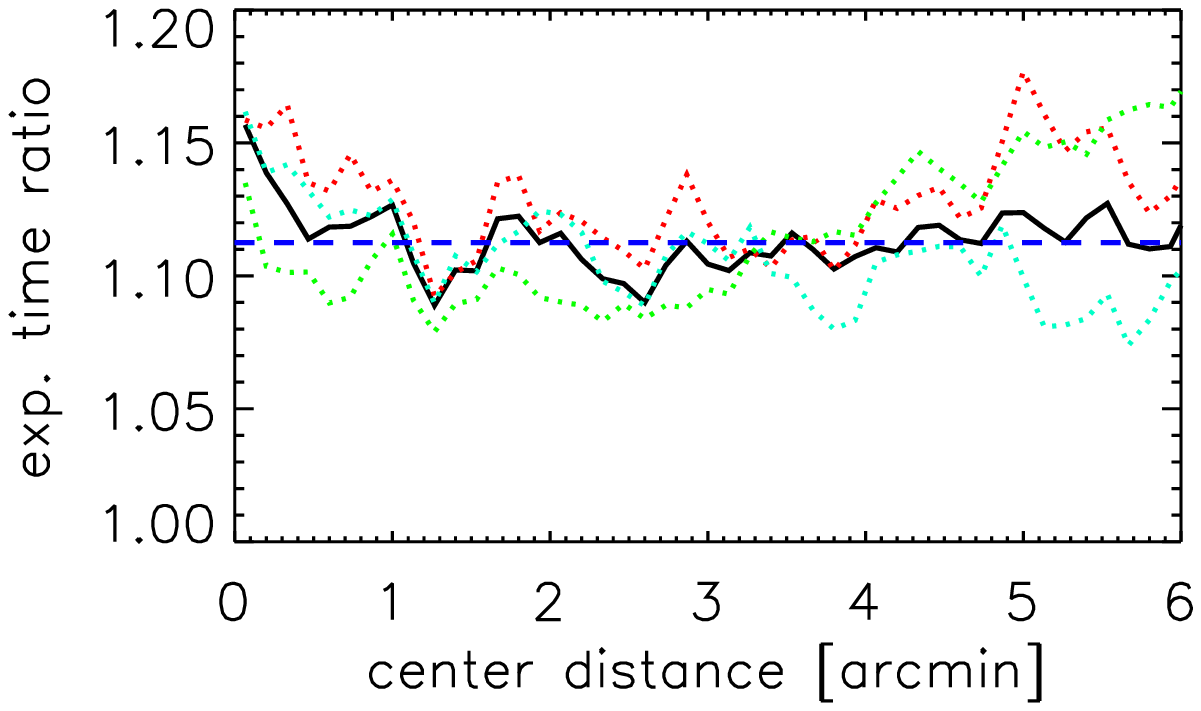}
%\vspace{-1ex}
\caption{Left: Comparison of the stacked effective exposure time radial profile %effective stacked radial exposure time profile
 with respect to the defined image center positions in the cluster fields (blue solid line), the average control field (black solid line), and the three individual control fields (dotted lines). Right: Fractional difference of the stacked effective exposure time profile
%effective stacked exposure times 
of the cluster field and the average control fields as a function of radial distance from the field centers (black solid line). The blue dashed line indicates the median exposure time difference of 11.3\%, dotted lines depict the individual control fields as before.
%The dotted lines shows the expected resulting fractional difference in limiting magnitude due to the slightly increased exposure time in the cluster field.
}
\label{f3a_Exp_Profile}       % Give a unique label
\end{figure}

A robust determination of the background number counts in the three control fields is of critical importance for the results of the stacking analysis. As discussed, some control field positions had to be placed at slightly larger off-axis angles compared to the cluster in order to avoid a significant overlap of the source lists at relatively small off-center distances (i.e.~$<$8\arcmin). Such a shift was applied to three of the 22 fields with up to 2\arcmin, to four fields with  2\arcmin-4\arcmin, and to two fields with $>$4\arcmin, which inevitably results in slight differences of the stacked effective clean exposure time in the cluster and control fields. Figure\,\ref{f3a_Exp_Profile} shows these stacked exposure time profiles (left panel), from which a median fractional exposure difference for the cluster field of dt$_{\mathrm{exp}}$/t$_{\mathrm{exp}}$$=0.113=$11.3\% is determined as displayed by the blue dashed line in the right panel. As an important cross-check, it can be confirmed that no significant systematic radial trend is present in the control field exposure time ratios (black line) with respect to the median. 

For the deep, background-limited XMM-{\it Newton} observations this translates into an average fractional difference of the flux limit of df$_{\mathrm{lim}}$/f$_{\mathrm{lim}}$$\simeq$$-$(dt$_{\mathrm{exp}}$/t$_{\mathrm{exp}}$)$^{0.5}$$\simeq$$-$3.4\%.   
%exposure map corrections
With the knowledge of the log$N$-log$S$ distribution from point source number counts in deep fields, the effect of the slightly increased sensitivity in the cluster field can be quantified. Using the measured faint end slope of the log$N$-log$S$ distribution of $\alpha\!=\!(1.65\pm0.05)$ \cite{Cappelluti2007a}, the fractional effect on the number counts can be determined as $dN/N\!\simeq\!-(\alpha-1)(dS/S)\!\simeq\!0.022\!\simeq\!2.2\%$. For a fair evaluation of excess counts in the cluster environment, the control field counts have thus to be scaled up by the small correction factor $b_{\mathrm{cor}}\!\simeq\!1.022$ that accounts for the slight average sensitivity differences.

\begin{figure}[b]
\centering
\includegraphics[angle=0,clip,width=0.68\textwidth]{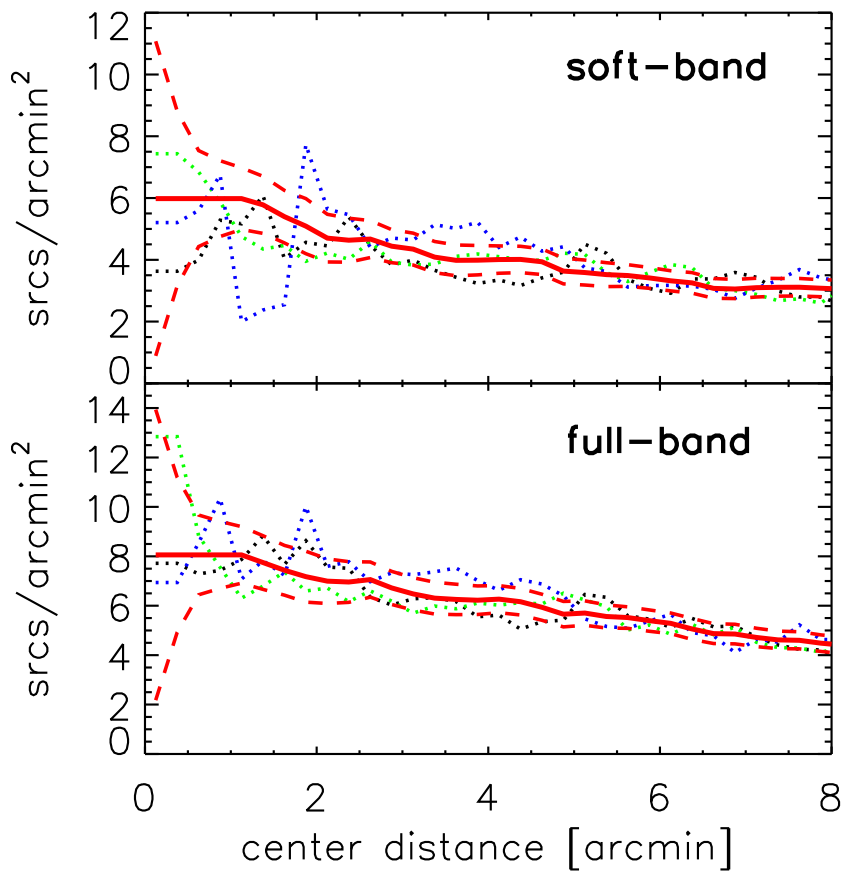}
%\includegraphics[angle=0,clip,width=0.68\textwidth]{Pub_v10_Smooth_BackgroundCounts_multi_RadialDistribution.ps}
%\vspace{-1ex}
\caption{Background model as a function of center distance for the detected sources  in the {\it soft-band} (top) and the  {\it full-band} (bottom).  The red solid line shows the average smoothed background model with 1\,$\sigma$ uncertainties indicated by the red dashed lines, whereas the colored dotted lines show the extracted counts in the three control fields. The gradual decline as a function of off-center angle reflects the radial change in effective exposure time as shown in Fig.\,\ref{f3a_Exp_Profile}.
}
\label{f3b_Back_Model}       % Give a unique label
\end{figure}

The sensitivity-corrected radial distribution of the background counts in the three control fields is displayed in Fig.\,\ref{f3b_Back_Model} (dotted lines) for the {\it soft-band} (top panel)  and {\it full-band} detection (bottom panel).  The final robust background model for the further analysis should on one hand reflect the gradual radial change in effective exposure time and on the other hand be a smooth function in order to suppress Poisson fluctuations in the individual radial bins, in particular close to the central position where the counts per bin drop to $\la$5.  This is achieved by applying a boxcar filter with a radius of four bins (each 15\arcsec) and inverse variance weighting to the averaged radial count distribution of the control fields. The resulting final average background models for the radial distribution in the {\it soft-band} and {\it full-band} are shown by the solid red lines in  Fig.\,\ref{f3b_Back_Model}, whereas the dashed red lines illustrate the estimated  1\,$\sigma$ Poisson  uncertainties.
\enlargethispage{2ex}

%.....................................................................................................................................................

%$\vspace{-1ex}
{ 
The determined background model may still be subject to residual systematic uncertainties not fully captured by the adopted approach with the three independent control fields and the applied average correction factor. Such residual systematic uncertainties include (i) slight effective exposure time offsets dependent on the exact locations of the optical axes of the different detectors, (ii) the effects of the removed extended and spurious sources, and (iii) slight geometric area mismatches at large off-center positions due to the non-axis-symmetric field-of-view edge. The magnitude of these potential systematic effects needs to be estimated for the combined stacked fields, which are the sums of the 22 random center positions across the XMM-{\it Newton} FoV.  
\enlargethispage{2ex}
The potential exposure time offsets (i) can be shown to amount to a negligible sub-per-cent effect.  
Similarly, the effect of the area covered by the removed extended sources (ii) is insignificant and stacking analysis tests without any sources removed showed the qualitative same results. The largest residual systematic arises from geometric edge effects (iii), which can be estimated 
analogous to the statistical spatial uncertainty of a one-dimensional random walk with 22 steps, since the edge effects are randomly positive or negative relative to the cluster field.  In this case,
 the step size for the random walk is given by the average  geometric mismatch of the analysis area per field of 8-9\,arcmin$^2$,  
 %per field given by the average  geometric mismatch of the analysis area per field. 
 resulting in a total statistical geometric mismatch after 22 steps of $\pm\sqrt{22}\times9\!\simeq\!\pm42$\,arcmin$^2$. 
 Using the average source densities yields approximate upper limits for the total impact of this effect of $\sigma_{\mathrm{sys}}$({\it full-band})$\la$12 sources and $\sigma_{\mathrm{sys}}$({\it soft-band})$\la$7 point sources. Consequently, the total systematic error budget is  estimated to be $\la$60\% of the total Poisson uncertainties derived in the next section and is hence still sub-dominant compared to the statistical errors.
}

\clearpage

%%%%%%%%%%%%%%%%%%%%%%%%%%%%%%%%%%%%%%%%%%%%%%%%%%%%%%%%
\section{~Results}
\label{s4_Results}

\noindent
In the following, the results of the X-ray stacking analysis are presented for the cumulative number counts, the radial distribution of excess counts, and an evaluation of source counts along the  principal cluster elongation axis. All results are given in stacked units,   i.e.~the sum of all 22 cluster environments is combined into a single radial X-ray source count profile. For all analyses a radial bin size for each independent ring segment of 15\arcsec \ is used and the maximal off-center radius is limited to 8\arcmin \ ($\sim$4\,Mpc) in order to avoid significant field-of-view edge effects. 

\vspace{-4ex}

%------------------------------------------------------------------------------------------------------------------------------------------
\subsection{~Cumulative Source Counts}
\label{s5_CumulativeCounts}

\begin{figure}[t]
\centering
\includegraphics[angle=0,clip,width=0.495\textwidth]{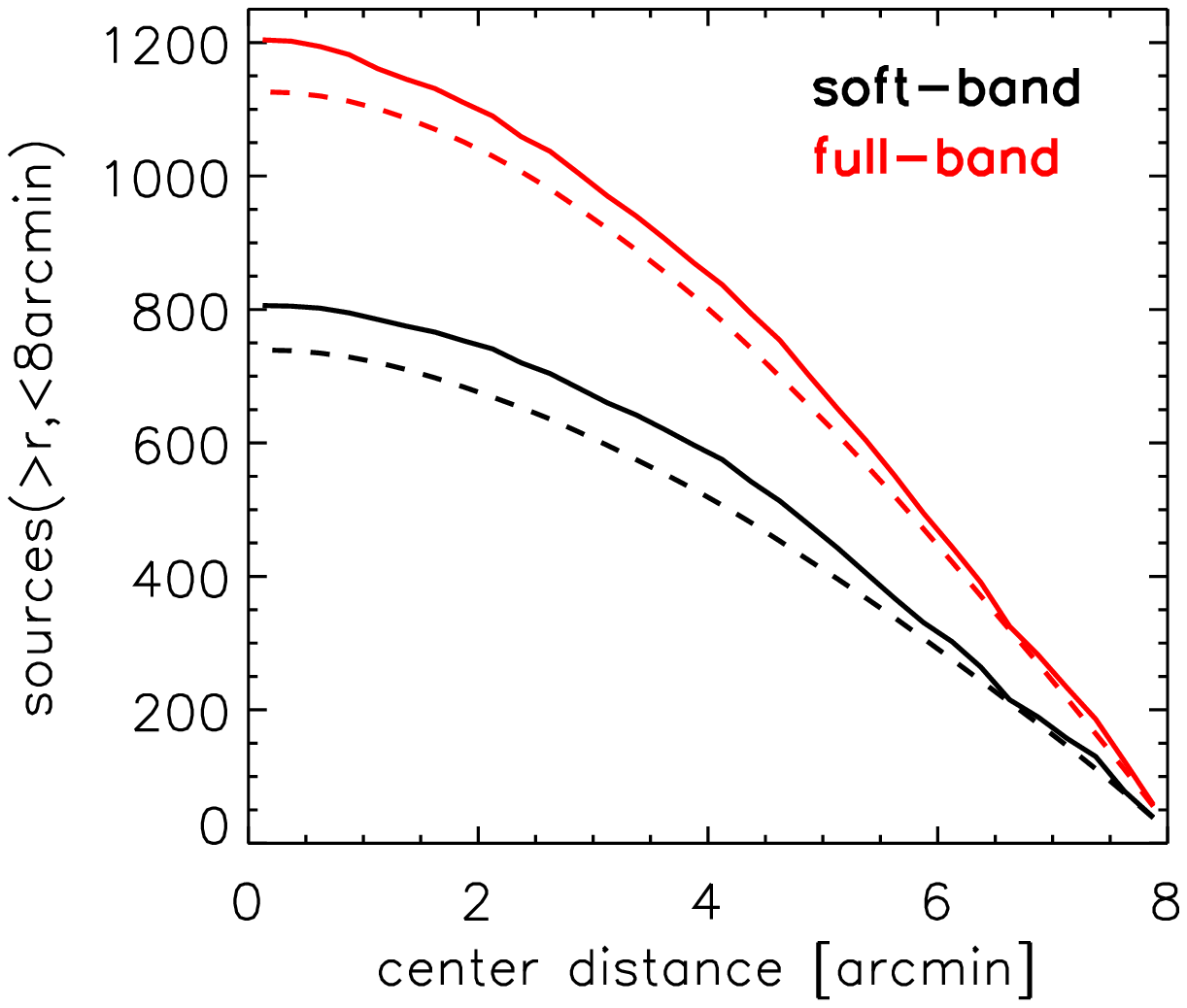}
\includegraphics[angle=0,clip,width=0.495\textwidth]{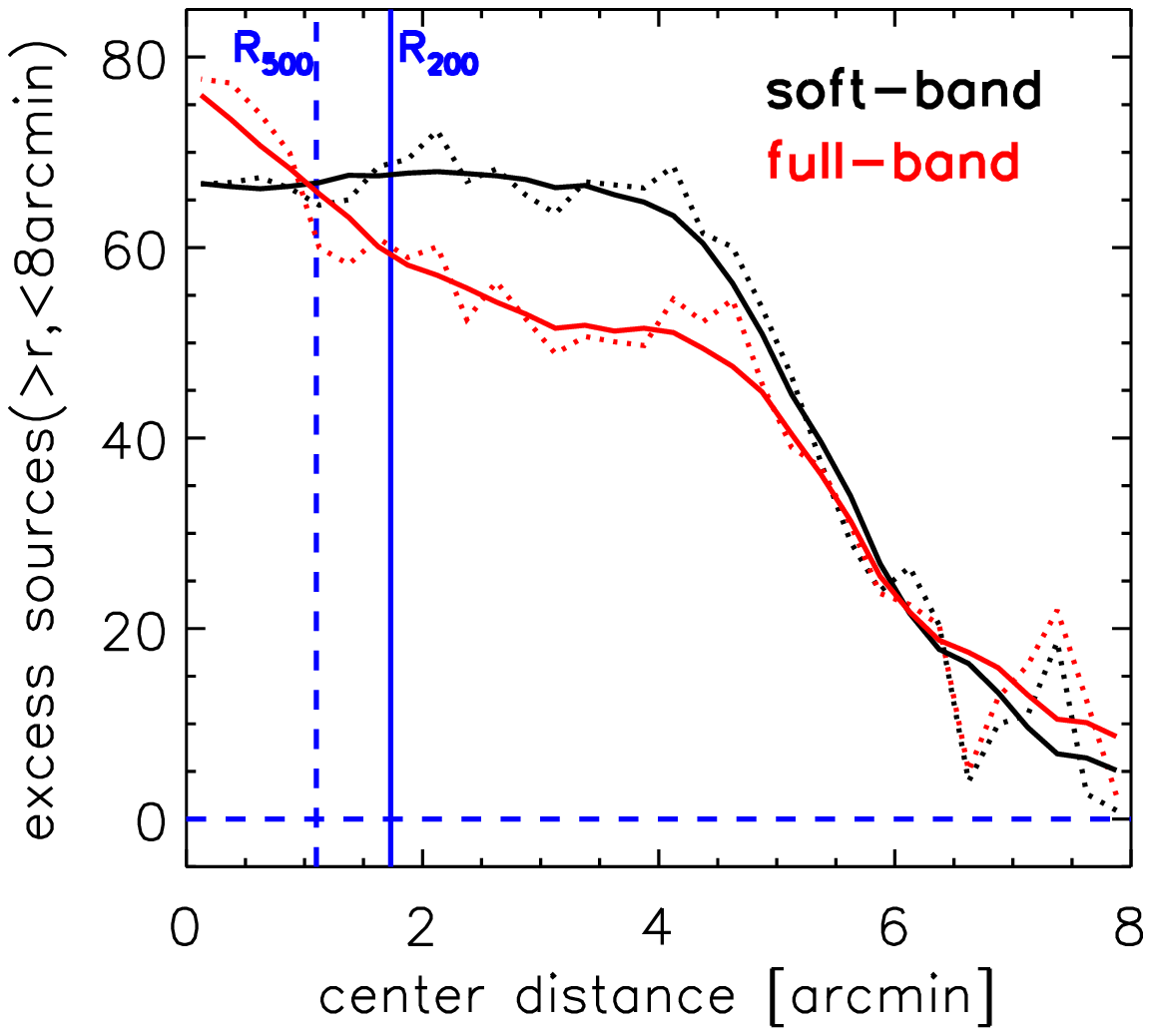}
%\includegraphics[angle=0,clip,width=0.49\textwidth]{Pub_v5_Cumulative_RadialDistribution.ps}
%\includegraphics[angle=0,clip,width=0.49\textwidth]{Pub_v3_DIFCumulative_RadialDistribution.ps}
%\vspace{-1ex}
\caption{Cumulative number count distributions starting from an off-center angle of 8\arcmin \ inwards.  Left panel: Total cumulative counts for cluster fields (solid lines) in comparison to
the median background models (dashed lines) for the {\it soft-band}  (black) and  {\it full-band}  detection (red).
% in comparison to the cluster field counts  displayed by the solid lines. 
Right panel: Cumulative distribution of excess counts after subtraction of the background model for the  {\it soft-band}  (black) and  {\it full-band} (red).  Dotted lines show the original measured distribution, whereas solid lines depict the smoothed trends after boxcar filtering. Average cluster radii R$_{500}$ (dashed) and R$_{200} (solid)$ for the distant cluster sample are marked by the blue vertical lines.}
\label{f4_Cumulative_Distrib}       % Give a unique label
\end{figure}

\noindent
The left panel of Fig.\,\ref{f4_Cumulative_Distrib} displays the cumulative number counts for the cluster (red and black) and average background field counts (blue) starting at a radial distance of 8\arcmin \ and moving inwards towards the center position. A significant excess of X-ray sources in the cluster field is apparent starting at off-center distances of about  6\arcmin \ in both the {\it soft-band}  (lower dashed lines)  and  {\it full-band}  counts (upper solid lines). 
For the {\it full-band} source list the number of excess X-ray counts  in the cluster field within the 8\arcmin \ analysis radius is about 78, while the 
{\it soft-band} shows approximately 67 excess counts. This translates into a fractional source excess in the distant cluster environments of +9.0\% in the  {\it soft-band}   and +6.9\% in the {\it full-band}.

Interestingly, the radial distribution of the excess counts for the two different band schemes is different, which is shown in the right panel of Fig.\,\ref{f4_Cumulative_Distrib}. The smoothed trends (solid lines) indicate that the main excess counts in the  {\it soft-band}  (black) originate at cluster-centric distance of 4-6\arcmin, corresponding to 2-3\,Mpc, while excess source counts in the  {\it full-band} rise sharply all the way to the center.

Applying Poisson statistics to the total cumulative background counts, which are based on three independent control fields, the statistical 1\,$\sigma$ uncertainties %errors 
can be estimated as $\sqrt{N_{\mathrm{tot}}/3}$, which amounts to $\sigma^{\mathrm{soft}}_{\mathrm{tot,stat}}\!\simeq\!15.7$ sources and  $\sigma^{\mathrm{full}}_{\mathrm{tot,stat}}\!\simeq\!19.4$ counts for the two different band schemes. 
The significance of the total detected X-ray source excess within a cluster-centric distance of  8\arcmin \ is hence 4.2\,$\sigma_{\mathrm{stat}}$ in the   
 {\it soft-band}  and 4.0\,$\sigma_{\mathrm{stat}}$  for the  {\it full-band} detection. The average %band-averaged 
 number of detected excess X-ray AGN per cluster environment  amount to $3.0\pm0.7$ {\it soft-band}  and $3.5\pm0.9$ {\it full-band} AGN, respectively, which are to be interpreted as lower limits since not the full geometric area out to  8\arcmin \ was covered in the detector FoV for each cluster environment.

 \begin{figure}[t]
\centering
\includegraphics[angle=0,clip,width=0.68\textwidth]{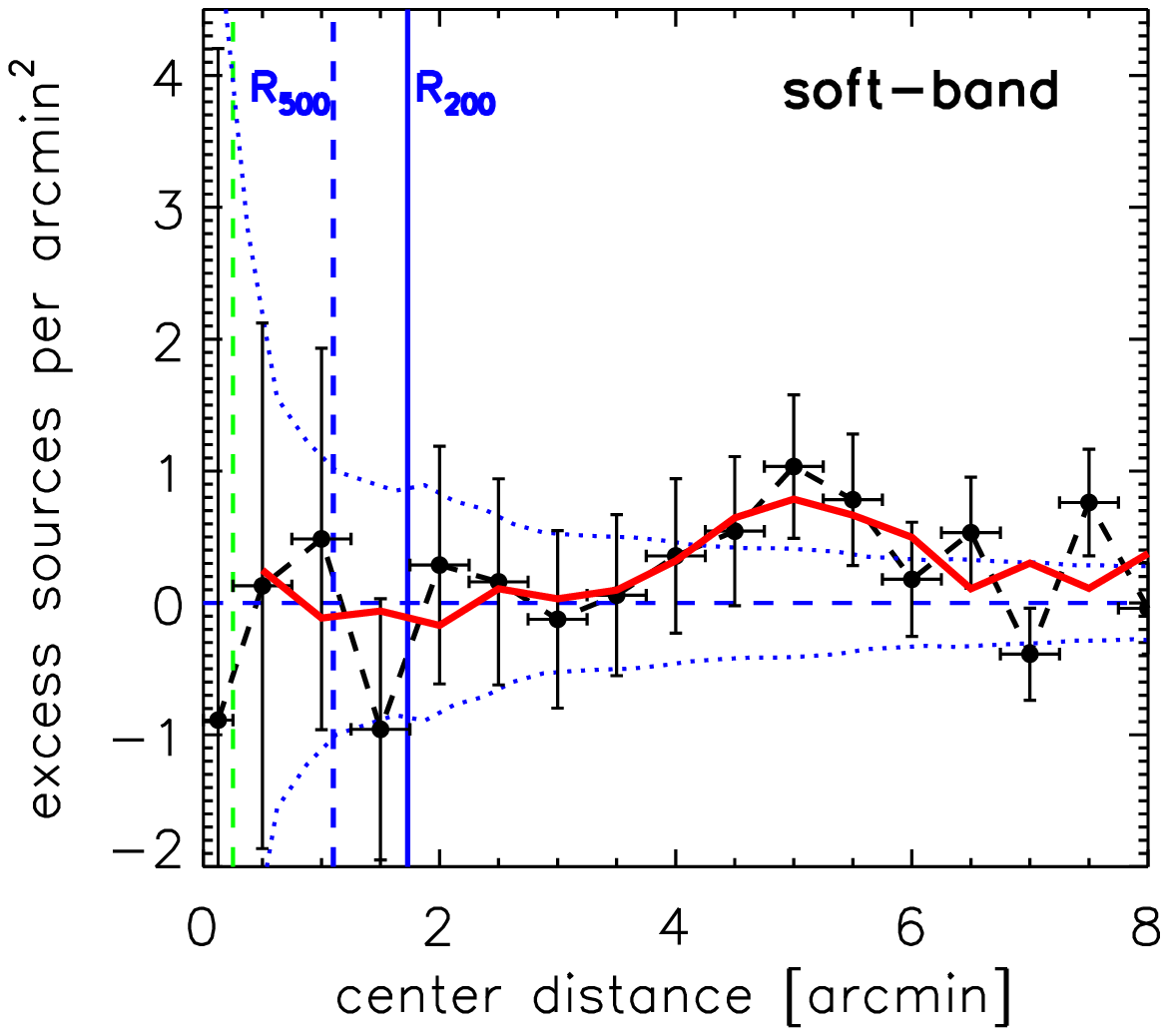}
\includegraphics[angle=0,clip,width=0.68\textwidth]{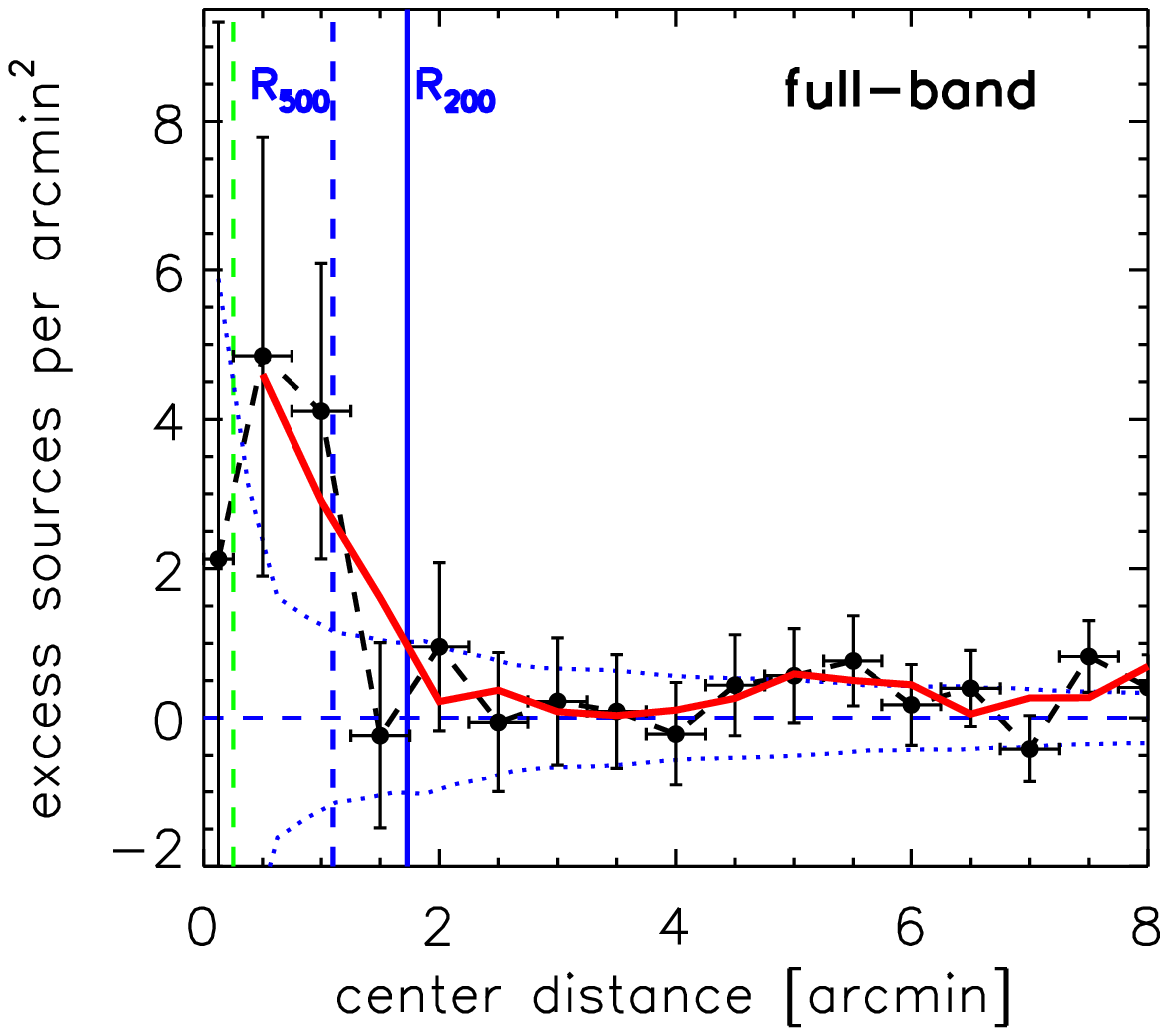}
%\includegraphics[angle=0,clip,width=0.495\textwidth]{v2a_Sm_DIF_035_24_RadialDistribution.ps}
%\includegraphics[angle=0,clip,width=0.495\textwidth]{v2c_ReBin_Sm_DIF_035_24_RadialDistribution.ps}
%\includegraphics[angle=0,clip,width=0.495\textwidth]{v2a_Sm_DIF_05_75_RadialDistribution.ps}
%\includegraphics[angle=0,clip,width=0.495\textwidth]{v2c_ReBin_Sm_DIF_05_75_RadialDistribution.ps}
%%\includegraphics[angle=0,clip,width=0.68\textwidth]{v2a_Sm_DIF_035_24_RadialDistribution.ps}
%%\includegraphics[angle=0,clip,width=0.68\textwidth]{v2a_Sm_DIF_05_75_RadialDistribution.ps}
%\includegraphics[angle=0,clip,width=0.68\textwidth]{v3_Sm_DIF_035_24_RadialDistribution.ps}
%\includegraphics[angle=0,clip,width=0.68\textwidth]{v3_Sm_DIF_05_75_RadialDistribution.ps}
%\vspace{-1ex}
\caption{Background subtracted radial distributions of excess X-ray sources for the {\it soft-band}  (upper panel) and  {\it full-band} (lower panel). Dotted blue lines indicate the 1\,$\sigma$ uncertainties about the mean background counts  (blue dashed). Data points and the black dashed line display the measurements in each of the independent %15\arcsec-wide 
radial bins, whereas the red solid line shows the boxcar smoothed radial trend. The green dashed vertical line on the left shows the 15\arcsec \ radius below which the measurements are biased due to the comparable spatial resolution of XMM-{\it Newton}.  The blue vertical lines mark the average cluster radii R$_{500}$ (dashed) and R$_{200} (solid)$. }
\label{f4_Radial_Distrib}       % Give a unique label
\end{figure}

\clearpage

%------------------------------------------------------------------------------------------------------------------------------------------
\subsection{~Radial Distribution of Excess AGN}
\label{s5_RadialAnalysis}

\noindent
The background subtracted radial distribution of excess X-ray sources in Fig.\,\ref{f4_Radial_Distrib} shows the data points for each individual bin with Poisson errors (black) and the smoothed radial trend (solid red line) for the {\it soft-band}  (upper panel) and  {\it full-band} detection (lower panel). 
The observed trends in the cumulative distribution of the previous section can now be evaluated for the independent data points with associated error bars and the given uncertainty in the subtracted background counts (blue dotted lines).  
For representation purposes (Figs.\,\ref{f4_Radial_Distrib}\,\&\,\ref{f5_Aligned_Distrib}) and better statistics in each bin, the displayed radial bins were increased to 30\arcsec \ each outside the biased central 15\arcsec-region (green dashed vertical line, see Sect.\,\ref{s2_SampleDescription}), which is also excluded for the smoothed solid trend lines.

In the  {\it soft-band}  (upper panel), the most significant feature in the radial distribution is  the hump between cluster-centric distance 4-6\arcmin \ as depicted by the red trend line. The covered solid angle at cluster-centric distances 4\arcmin$\le$r$\le$6\arcmin \ is 63 square arcmin implying that significant excess counts normalized to a unit area are hard to measure in comparison to the background counts. However, in this 4\arcmin$\la$r$\la$6\arcmin \  range six independent adjacent radial bins show an excess, of which four have more than 1\,$\sigma$ significance each, yielding a high combined confidence that this feature is indeed real. The hump is also discernible in the  
{\it full-band} detection (lower panel), although at lower statistical significance.

The most prominent  feature in the   {\it full-band} detection is the strong rise of excess X-ray sources at radial distances below 2\arcmin, which correspond approximately to the radius inside the fiducial R$_{200}$ (vertical blue line) of the distant cluster sample\footnote{{R$_{200}$ (R$_{500}$) are the radii for which the mean enclosed total mass density of the cluster is 200 (500) times the critical
energy density of the Universe $\rho_{\mathrm{cr}}$(z) at the given redshift $z$. The considered distant cluster sample has an average R$_{200}$ (R$_{500}$) of about 100\arcsec \ (64\arcsec) with a standard deviation of $\pm$20\%.}
}. 
Owing to the relatively small enclosed central area with a corresponding low number of total counts, the statistical uncertainties inside the fiducial cluster regions are quite significant. Nevertheless, the two innermost bins outside the core region indicate a clear and significant trend of a steep inner radial profile of  {\it full-band} detected excess sources. The inner profile of the {\it soft-band}  sources in the upper panel, on the other hand, is fully consistent with a null excess of sources inside radii r$<$2\arcmin.

%\vspace{-1ex}
%------------------------------------------------------------------------------------------------------------------------------------------
\subsection{~X-ray Counts along the Principal Cluster Elongation  Axis}
\label{s5_AlignmentAnalysis}

\begin{figure}[t]
\centering
\includegraphics[angle=0,clip,width=0.68\textwidth]{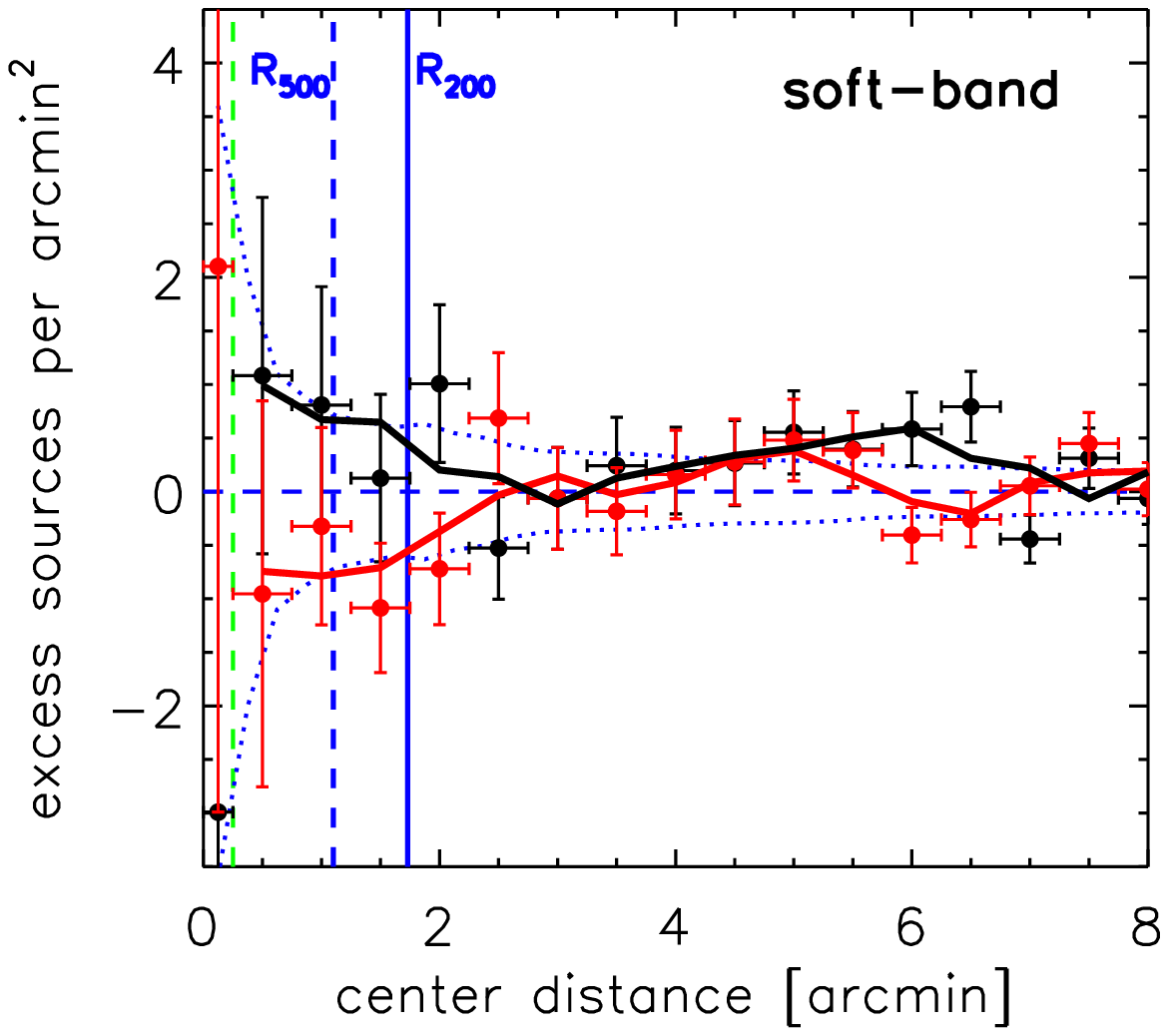}
\includegraphics[angle=0,clip,width=0.68\textwidth]{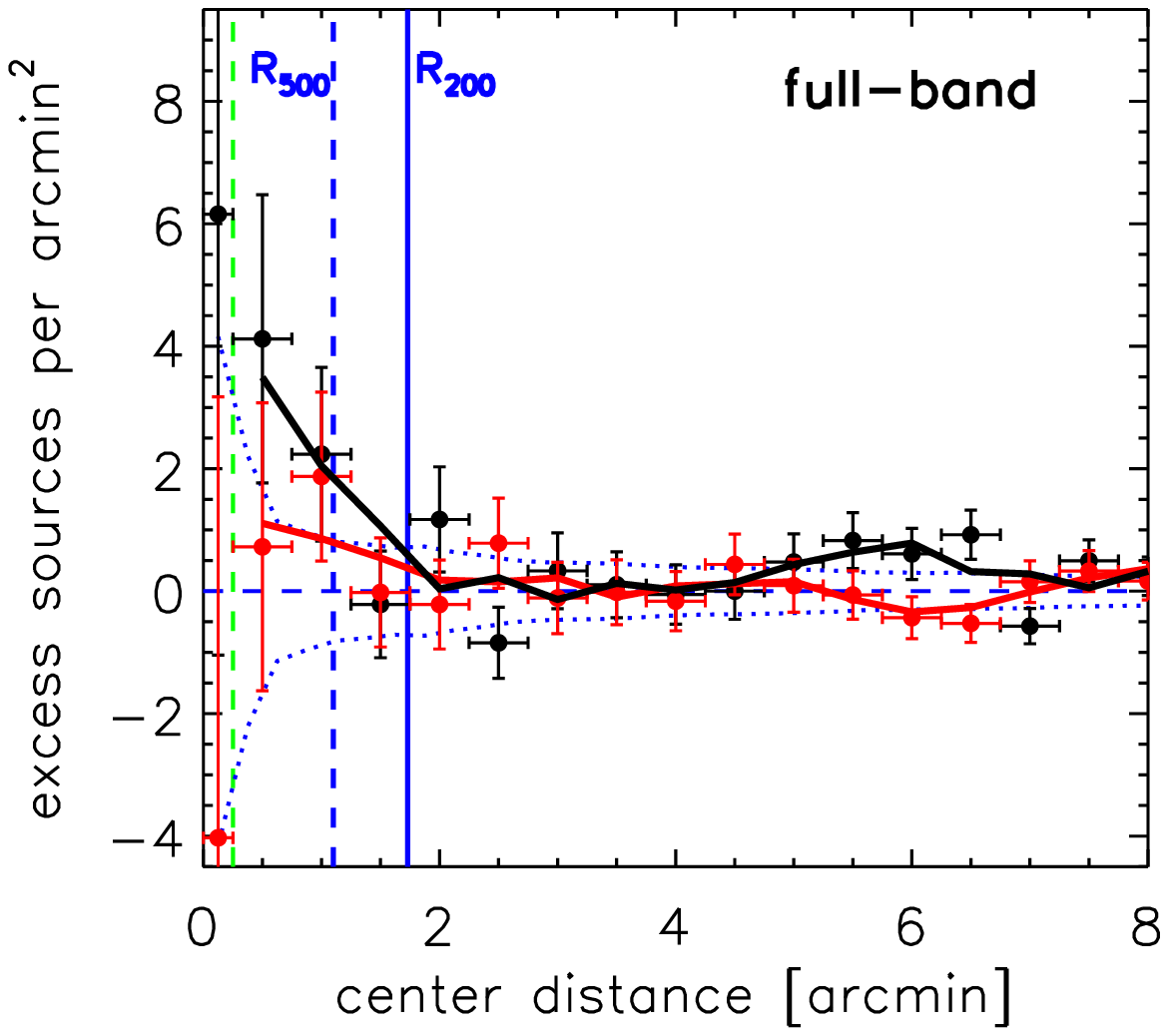}
%\includegraphics[angle=0,clip,width=0.68\textwidth]{v1_PubROT_DIF_035_24_RadialDistribution.ps}
%\includegraphics[angle=0,clip,width=0.68\textwidth]{v1_PubROT_DIF_05_75_RadialDistribution.ps}
%\vspace{-1ex}
\caption{Same as Fig.\,\ref{f4_Radial_Distrib} but now distinguishing the sectors aligned with the cluster elongations (black points and solid black line) and the perpendicular direction (red points and solid red line).}
\label{f5_Aligned_Distrib}       % Give a unique label
\end{figure}

\noindent
In order to investigate the possibility that AGN activity occurs along a preferential direction in the cluster-frame system, the radial profile analysis for excess X-ray counts was repeated with the rotated stacked source lists that aligned the principal elongation axis of the cluster X-ray emission in the vertical direction (see Sect.\,\ref{s2_StackinAna} and bottom panels of Fig.\,\ref{f2_Extraction_Cluster}).   
The radial distribution was then split into two disjoint sectors, the  {\it aligned} sector along the cluster elongation direction %axis  
covering the range within an angle of $\pm$45\degr \ to the vertical axis with the cluster center as origin, and the  {\it perpendicular} sector within $\pm$45\degr \ to the horizontal coordinate axis. 
The average background models (Fig.\,\ref{f3b_Back_Model}) and the associated uncertainties were adopted to account for the factor of two smaller total area covered by each of the two sectors. 
The background subtracted radial counts along the   {\it aligned} (black)  and  {\it perpendicular}  sectors (red) are displayed in Fig.\,\ref{f5_Aligned_Distrib} for the   {\it soft-band}  (upper panel) and  {\it full-band} detection (lower panel).

For the {\it full-band} counts in the lower panel 
the  profiles at r$<$2\arcmin \ indicate that the X-ray source excess inside the fiducial  cluster radius R$_{200}$ 
is dominated by AGN along the  cluster-{\it aligned}  direction (black line). Additionally, the secondary hump at 4.5\arcmin-6.5\arcmin \ is now more significant in the {\it aligned} (black) direction, while consistent with a null excess for the  {\it perpendicular}  sector (red).
However,  owing to the lower number of excess counts in the individual split sector bins with a corresponding decreased statistical significance  these results are to be considered as tentative. 

%Beyond a radius of 2\arcmin \ no significant features are currently discernible. 
%However,  the lower statistical significance in the individual sector bins  
%and a potential bias that cluster embedded faint AGN are partially responsible for the observed cluster elongation direction make this only a tentative result. 

For the   {\it soft-band}  detection in the upper panel, the 4\arcmin-6\arcmin \ hump appears also to be enhanced in the   {\it aligned} (black) direction around r$\sim$6\arcmin.
%is still present in both sectors. 
Below a radius of 
2\arcmin \  the sector counts seem to split into an excess in the  {\it aligned}  %direction 
and a deficit in the  {\it perpendicular} sector, but both are on the level of the background uncertainty. %, i.e.~of tentative nature.

\clearpage

%\vspace{-1ex}
%------------------------------------------------------------------------------------------------------------------------------------------
\subsection{~Robustness of Results}
\label{s5_Robustness}

\noindent
The tentative nature of the latter results in Sect.\,\ref{s5_AlignmentAnalysis}  with the separated sectors along two directions illustrate the limits of the statistical detectability of radial features in the X-ray source distribution for the currently available sample size of distant X-ray clusters. However, the results based on the overall excess counts in the radial distributions in Sect.\,\ref{s5_CumulativeCounts}\,\&\,\ref{s5_RadialAnalysis} are detected on the 4\,$\sigma$ level {(for r$<$8\arcmin)} and can hence  be considered as robust.

%\clearpage

In particular the central rise of the {\it full-band}  counts and the  {\it soft-band}  4\arcmin-6\arcmin \ hump seem to be real features in the radial distribution of distant cluster environments with an average of 1-2 excess sources per cluster in each feature. 
A central rise in the {\it full-band} while flat in the  {\it soft-band} could well be physically explained by a dominance of faint and spectrally hard cluster AGN. With on average about one {\it full-band} excess source per cluster in the inner $\sim$2\arcmin, the central part of the background model would have to be underestimated by almost a factor of two to attribute the observed cumulative excess to a background systematic. 

{ 
The cumulative counts for the  4\arcmin-6\arcmin \ hump in the  {\it soft-band} exceed the  {\it full-band} counts by about 15 sources around r$\simeq$4\arcmin \ (see right panel of Fig.\,\ref{f4_Cumulative_Distrib}). On the other hand, all  {\it soft-band}  sources should also be part of the  {\it full-band} catalogs, i.e.~the cumulative {\it full-band} excess would be expected to be at least as high for the {\it soft-band}. However, this slight discrepancy at intermediate off-center angles between 2-5\arcmin \ in the cumulative excess distribution   
%, on the other hand, 
can likely be attributed to  Poisson noise %uncertainty
 in the {\it full-band} background counts, which was estimated to be slightly higher in %of the same order of 
 magnitude compared to the measured difference of %about 15
  excess   {\it soft-band}  sources at the hump location. The  {\it full-band}  signal at these intermediate off-center distances hence seems to be diluted by the larger Poisson noise at a level within the statistical expectations. 
  }
 
 %Since the {\it full-band}  encompasses the  {\it soft-band},  basically all objects of the latter source list  can be expected to have a counterpart in the former, and excess counts should hence be at least on the same level. 
% However, 
The presence of the hump in the  {\it full-band} itself, although at lower confidence level, provides some extra confidence in its real nature. 
Other cross-checks with the exposure time profile (Fig.\,\ref{f3a_Exp_Profile}) and the background (Fig.\,\ref{f3b_Back_Model}) do not reveal any particular radial feature in the  4\arcmin-6\arcmin  \ range that could attribute the observed hump to a systematic.   
%Furthermore, any reasonable systematic uncertainty related to the applied global 2.2\% correction to the background count model would be small compared to the Poisson errors and can hence be neglected. 
{
The combined systematic error budget was estimated in Sect.\,\ref{s2_StackinAna} to be at most 60\% of the total statistical Poisson error ($\la$12 sources), which is far below the observed signal.
}
The cumulative excess magnitude in the hump of more than three dozen sources also %rules out 
makes the possibility of
individual systematic structures at $\sim$5\arcmin \  in the cluster fields an unlikely explanation.

%difference in the shoulder of cumulative excess counts of 15 to be attributed to Poisson fluctuation.
%cross-comparison between {\it soft-band}  and  {\it full-band}
%smooth exposure time curves at 4-6 amin
%systematic structure, e.g. a cluster environment at 5arcmin distance (few cases)

%%%%%%%%%%%%%%%%%%%%%%%%%%%%%%%%%%%%%%%%%%%%%%%%%%%%%%%%
\section{~Discussion}
\label{s5_Discussion}
\noindent

%------------------------------------------------------------------------------------------------------------------------------------------
\subsection{~Spectroscopically Confirmed AGN}
\label{s5_SpecAGN}

\noindent
The ultimate proof to quantify the statistical properties of AGN in high-z  clusters  from the centers to the large-scale environments  would require a combination of high-resolution X-ray imaging plus an optical spectroscopic identification to identify all X-ray point sources. Such a data set is currently not yet available for sizable samples of  $z\!>\!0.9$ X-ray luminous systems.
However, as a starting point the observed X-ray point source catalogs  can be cross-matched with public redshift  data from the NASA Extragalactic Data Base\footnote{\url{http://nedwww.ipac.caltech.edu}} (NED) complemented by individual cluster publications \cite{Hilton2010a}.
The resulting list of 15 spectroscopically confirmed and X-ray identified AGN within 12\arcmin \ of the cluster positions and spectroscopic rest-frame velocity offsets of less than $\pm$3000\,km/s is shown in Table\,\ref{tab_Spec_AGN}.

Five of the spectroscopic AGN are located at  cluster-centric angular separations  8\arcmin$<$r$<$12\arcmin, corresponding to projected distances of 4-6\,Mpc. Ten spectroscopic members are found at the probed radial distance range of $\le$8\arcmin, four of them at $<$4\arcmin \ ($<$2\,Mpc) and six in the range 4\arcmin$<$r$<$8\arcmin (2-4\,Mpc). The two objects A04 and A05 at the center of XMMXCS\,J2215.9-1738 at $z\!=\!1.46$  are not spatially resolved with  XMM-{\it Newton} \cite{Hilton2010a}, all others are identified X-ray point sources\footnote{Except object A12, which is outside the XMM-{\it Newton} FoV, but is classified in NED as QSO.}.

Out of the measured {\it full-band} %({\it soft-band})
X-ray point source excess of about 78 ($\pm$25\%) %(67$\pm$25\%)  
in the cluster environments at r$\le$8\arcmin, eight or about 10\% are %hence
 spectroscopically confirmed X-ray AGN at the cluster redshift.
As can be expected, the serendipitous spectroscopic identification rate of cluster AGN in this high-$z$ sample is still fairly low, but sufficiently high to confirm the presence of X-ray AGN in the large-scale structure environment  of the probed systems.

%\pagebreak

%spectroscopic AGN, with visually confirmed XMM X-ray emission (softband) 

%A12 without direct X-ray ID, but listed as QSO

% ------------------------------- TABLE with Cluster Sample --------------------------------------------------------------
% 
%:SPECAGN_TABLE
\begin{table*}[t]
%\fulltable
\caption{ List of spectroscopically confirmed X-ray AGN within a search radius of 12\arcmin \ ($\simeq$6\,Mpc) from the cluster centers and with a spectroscopic redshift that is within $\pm$3000\,km/s of the cluster restframe velocity. The table lists the cluster environment ID of Table\,\ref{tab_Cluster_Sample} in column (1), an AGN ID in (2), the angular distance to the cluster center in (3), the AGN redshift in (4), the RA and DEC coordinates in (5-6), the source name as listed in NED in (7), and a redshift reference in (8). 
}\label{tab_Spec_AGN}
\centering
\begin{tabular}{ c c c c c c l l}
\hline \hline     
Cl. Env. &	AGN ID	& Distance	& $z$	&	RA	&		DEC	&		NED Name &		Reference	 \\  
 &		& arcmin	& 	&	J2000	&	J2000	&		 &	 \\  
 (1)  &  (2) &  (3)  & (4)  & (5)  & (6)  & (7)  & (8)     \\	
 \hline	
C03	&		A01&	9.52&	1.486&	03:39:08.3&	+00:29:25&	SDSS J033908.27+002924.9 & \cite{Schneider2007a}\\    
C03	&		A02&	11.3&	1.483&	03:39:12.3&	+00:30:56&	SDSS J033912.34+003055.5& \cite{Schneider2007a} \\ 
C04	&		A03&	1.31&	1.462&	22:16:03.7&	-17:38:31&   XMMU J221603.6-173830 & \cite{Carrera2007a}\\ 
C04	&		A04&	0.24&	1.462&	22:15:59.1&	-17:37:54&	PS1 & \cite{Hilton2010a} \\ 
C04	&		A05&	0.12& 	1.453&	22:15:58.9&	-17:38:10&	PS2 & \cite{Hilton2010a} \\ 
C06	&		A06&	6.36&	1.373&	15:32:06.0&	-08:30:55&	XMS J153206.0-083055& \cite{Barcons2007a}\\ 
C07	&		A07&	5.32&	1.330&	00:36:18.6&	-43:13:20&	IRAC 109477& \cite{Feruglio2008a}\\ 
C07	&		A08&	6.52&	1.318&	00:35:23.9&	-43:16:38&	IRAC 229193      &  \cite{Sacchi2009a}\\ 
C07	&		A09&	9.02&	1.334&	00:36:22.1&	-43:19:03&	IRAC 111580& \cite{Sacchi2009a}\\ 
C08	&		A10&	7.81&	1.234&	12:53:08.3&	-29:20:10&	GMOS-F2 08 & \cite{Tanaka2009a} \\ 
C10	&		A11&	3.08&	1.179&	03:02:14.8&	+00:01:25&	SDSS J030214.82+000125.3& \cite{Schneider2007a}\\ 
C12	&		A12&	11.6&	1.110&	22:06:26.0&	-01:52:01&	FBQS J2206-0152& \cite{Veron2001a}\\ 
C13	&		A13&	10.8&	1.120&	03:38:10.2& +00:23:25&	SDSS J033810.16+002325.1&\cite{Schneider2007a} \\ 
C15	&		A14&	7.97&	1.053&	02:26:38.0&	-04:19:45&	VVDS 020465089& \cite{LeFevre2005a}\\ 
C21	&		A15&	7.66&	0.932&	01:04:02.8&	-06:36:00&	XMMU J0104.1-0635 000 & \cite{Barcons2002a}\\ 

\hline
\end{tabular}
\end{table*}

%\clearpage

%------------------------------------------------------------------------------------------------------------------------------------------
\subsection{~Selection Effects and Comparison to Chandra Observations}
\label{s5_SelectionEffects}

\noindent
{The considered distant galaxy cluster sample is X-ray selected based on XMM-{\it Newton} data as discussed in Sects.\,\ref{s2_SampleDescription}\,\&\,\ref{s2_XrayData}, implying that potential selection effects could have an impact on some of the results presented in Sect.\,\ref{s4_Results}.
All clusters were originally detected as {\em extended} X-ray sources, with most of the detection weight originating from the soft band owing to the expected spectral properties of high-$z$ thermal ICM emission. In principle, cluster-embedded X-ray point sources could have both an  (negative) anti-bias or (positive) bias effect on the  XMM-{\it Newton}  selection. An anti-bias, i.e.~a missed fraction of proper cluster sources, arises when a bright central AGN is dominating over the extended thermal emission in the soft band, resulting in a potential misclassification of the cluster as an X-ray point source. On the other hand, an opposite bias effect could arise in the cases where a weak unresolved cluster AGN adds a subdominant fraction of flux to the underlying extended cluster emission or where the superposition of multiple embedded X-ray point sources even mimic an X-ray extend at the given XMM-{\it Newton} resolution limit.  

%Possible selection effects: anti-bias etc
%vs optical selection
Neither of the described possible anti-bias or bias effects are currently accessible to a robust quantitative evaluation since this would require large well-defined $z\!>\!0.9$ cluster samples based on different selection techniques (e.g. IR, SZE, and X-ray) and followed-up with high-resolution  {\it Chandra} observations and extensive optical spectroscopy  that are not yet available for such a study.
The two most distant spectroscopically confirmed infrared selected systems in the galaxy group mass regime, CL\,J1449+0856 at $z\!=\!2.07$ \cite{Gobat2011a} and SXDF-XCL\,J0218-0510 at $z\!=\!1.62$ \cite{Pierre2011a}, feature both a central X-ray point source that dominates the X-ray emission and could hint at a common occurrence of central AGN in low-mass systems  at $z\!\ga\!1.6$.
However, given the different system masses, redshift regime, and selection techniques such first hints based on low-number statistics may not be representative for the considered X-ray selected high-$z$ cluster sample. 

A meaningful cross-check on the influence of potentially unresolved embedded cluster AGN for the cluster sample of this work can currently only be obtained from the available published {\it Chandra} observations of a subsample of the five systems C04 \cite{Hilton2010a}, C05 \cite{Rosati2009a}, C08  \cite{Rosati2004a}, C16 \cite{Maughan2008a}, and C19  \cite{Fassbender2011b} listed in Table\,\ref{tab_Cluster_Sample}. For the four systems at $z\!\la\!1.4$ (C05, C08, C16, C19)   {\it Chandra} did not reveal any additional central point sources embedded in the extended ICM emission. Only for the system C04 at $z\!=\!1.46$ the high-resolution observations disclosed the two additional sub-dominant cluster AGN A04 and A05 (see Table\,\ref{tab_Spec_AGN} and Sect.\,\ref{s5_SpecAGN}) at r$\la$15\arcsec  \ that are unresolved with XMM-{\it Newton} both for the soft- and full-band detection. 
The cross-check with this {\it Chandra} sub-sample  is hence consistent with a fair XMM-{\it Newton}  accounting of X-ray point sources in the distant cluster environments outside the biased central 15\arcsec  \ radius, which was excluded from the trend analysis of the radial profiles in Sects.\,\ref{s5_RadialAnalysis}\,\&\,\ref{s5_AlignmentAnalysis}.  In particular, there is no indication that the soft-band radial point source profile could be anti-biased in the range $0.25\arcmin\!<\!r\!\la\!2\arcmin$ due to the presence of extended ICM emission. Similarly, the {\it Chandra} sub-sample does not reveal any cases of systems that entered the distant X-ray cluster sample solely based on a bias effect of embedded point sources.

In summary, although some sample selection effects inherit to the XMM-{\it Newton} discovery method for the distant clusters may apply compared to an ideal (non-existent) sample, a qualitative cross-check with the available {\it Chandra} observations for five systems did not reveal any evident biases concerning the radial profile of point sources in the main targeted cluster-centric distance range of    $0.25\arcmin\!<\!r\!\la\!8\arcmin$.

%Chandra observations: 5 published observations
%Conclusions

}

%:CURRENT_WORK

%------------------------------------------------------------------------------------------------------------------------------------------
\subsection{~Comparison to Previous Studies}
\label{s5_ResultComparison}

\noindent
The high-$z$ results of this work are consistent with numerous previous  {\it Chandra}  studies that find a general trend of increasing AGN activity in cluster environments with redshift for X-ray selected samples  \cite[e.g.][]{Martini2009a,Branchesi2007a,Eastman2007a,Cappelluti2005a} and infrared-selected clusters \cite{Galametz2009a}. 
%a) general trend
%IR selected clusters
Although the used sample is X-ray selected with XMM-{\it Newton}, implying the discussed natural anti-bias against potential systems with bright central soft-band AGN,
%\footnote{A bright central soft-band point source hampers the detection of any underlying extended X-ray emission and hence decreases the chance of being included in the sample.}, 
a centrally rising radial distribution of lower luminosity AGN was found with about one detected excess source per system within a projected radius of 1\,Mpc.  However,  because of the limited spatial resolution of XMM-{\it Newton}, the recovered excess sources are incomplete in the very core (r$\la$15-20\arcsec), where individual systems at the highest  accessible redshifts
have revealed embedded cluster AGN as discussed in the previous Sect.\,\ref{s5_SelectionEffects}.
%.\cite{Hilton2010a,Pierre2011a}.

%b) central peak 
%b2) incomplete in center
%bias against systems with central high-luminosity AGN

%At the presently highest accessible redshifts

%3 spectroscopically identified AGN in the environment of the infrared selected cluster IRC-0218A at $z\!=\!1.62$ 
%including a central point source 
%\cite{Pierre2011a}

Concerning the detected secondary  4\arcmin-6\arcmin-hump (2-3\,Mpc), \citet{Ruderman2005a} reported on the same general radial profile shape of the X-ray point source excess  around very massive MACS clusters at intermediate redshifts ($0.3\!<\!z\!<\!0.7$) with a central spike plus an additional broad excess at projected cluster-centric distances of 2-3\,Mpc. They interpreted this finding as evidence for distinct triggering mechanisms of nuclear activity in the center through close encounters of infalling galaxies, and the cluster-field interface, where galaxy  mergers fuel the central super-massive black holes.    
On the other hand,   \citet{Gilmour2009a} did not recover the secondary radial excess feature in their larger sample of 148 clusters at $z\!<\!0.9$. {Their}
tests with cluster subsamples revealed that a point source excess at r$>$2\,Mpc  may be systematically boosted by structures related to non-associated foreground clusters in the FoV   \cite{Gilmour2009a}. 

With respect to such a test, the current sample size of 22 high-$z$ cluster fields is still too small to discard all fields with a potential contamination from foreground structures, which are ubiquitous at the exposure depth of the considered observations.  For the nature of the 4\arcmin-6\arcmin-hump, a systematic bias related to foreground structures can hence not be fully ruled out at this point, although the measured excess in this region would require multiple such structures not captured by the background model.  A test with the two most crowded fields removed from the analysis (excluding clusters C03, C04, C09, C13, C17, and C22) showed no qualitative difference concerning  the shape and the presence of the hump.  %., but only in the normlization   

On the other hand, a cluster-associated excess of X-ray AGN at projected cluster-centric radii of r$>$4\arcmin \ ($>$2\,Mpc) must be present as revealed  by the 11 spectroscopically confirmed objects in Table\,\ref{tab_Spec_AGN} in the range 4\arcmin$<$r$<$12\arcmin. Moreover, \citet{Rumbaugh2011a} showed in their spectroscopic study of cluster environments in the redshift range 0.7-0.9 that half of the AGN host galaxies are located at projected distances of $>$1.5\,Mpc away from the nearest cluster or group. In particular, they also confirmed a significant difference in AGN X-ray luminosity between  objects  in the dense inner cluster regions and the low-density large-scale structure environments in the sense that the central objects show low luminosities, while the brightest objects are all found at large cluster centric distances. 

%c) secondary peak

%Gilmore
%no
%
%spec results: 11 at r$>$4\arcmin
%not possible to separate sample, not enough statistics

%------------------------------------------------------------------------------------------------------------------------------------------
\subsection{~The Distant Cluster-AGN Connection}
\label{s5_Cluster_AGNconnection}

\noindent
Taking the presented results at face value, the following scenario for AGN activity in distant X-ray cluster environments at $0.9\!<\!z\!\la\!1.6$ emerges. The detected significant (4\,$\sigma$) excess of X-ray AGN in the distant cluster environments at angular radii of up to 8\arcmin \ is split into two distinct populations that reflect different triggering mechanisms of nuclear activity. 

At  cluster-centric radii less than the fiducial average cluster size of the sample of R$_{200}$$\simeq$830\,kpc$\simeq$1.7\arcmin \ 
a centrally peaked population of low-luminosity X-ray AGN exists. The conclusion of low X-ray luminosities (or a heavily absorbed type-II spectrum) for this population originates from the fact that the central excess is not  observed in the {\it soft-band} detection with its higher source significance threshold. The tentative result of Sect.\,\ref{s5_AlignmentAnalysis} suggests that the AGN activity preferentially occurs along the main matter infall axis of the cluster, as indicated by the principal elongation axis of the extended cluster emission. This nuclear activity is likely triggered by close encounters of infalling objects and is mostly found in  red or green transitional galaxies as reported in \citet{Rumbaugh2011a}. Overall, the detected (incomplete) excess of about one AGN source per cluster field inside its fiducial radius R$_{200}$ is still very moderate when considering the expected \cite[e.g.][]{Giodini2009a} average enclosed stellar mass of $\simeq\!4\!\times\!10^{12}$\,M$_{\sun}$ for the cluster systems. Comparing this value to the average AGN abundance in the COSMOS field in the same $0.9\!\le\!z\!\le\!1.6$ redshift range  of one X-ray AGN per $2\!\times\!10^{12}$\,M$_{\sun}$ in stellar mass (Bongiorno et al., in prep.) suggests that the  AGN activity in high-$z$ cluster environments  is still  suppressed by a factor of two compared to the field abundance.

%., suggesting that AGN activity in high-$z$ clusters is still suppressed by the environmental influence compared to the field.

The second AGN population is preferentially located at cluster-centric distances of 2-3\,Mpc as part of the observed  4\arcmin-6\arcmin-hump of excess sources. The  {\it soft-band}  detection of this feature indicates significantly higher soft-band X-ray luminosities compared to the central AGN. The projected distance of the hump location corresponds to about 3$\times$R$_{200}$
of the average cluster radius. At this distance, the relative galaxy velocities are still small while the object density is already significantly enhanced, resulting in an environment where major merging processes are expected to be efficient. 
The observed AGN excess in the larger-scale environment of distant X-ray clusters is hence likely to be attributed to merger-induced nuclear activity in `quasar mode'  \cite{Croton2006a} {occurring at a potential sweet spot for merging events in the infall regions at cluster-centric projected distances of around  3$\times$R$_{200}$}.

%a) relate to average R200 of sample cluster

%b) central triggering, central peak, and still moderate values

%c) secondary peak

%------------------------------------------------------------------------------------------------------------------------------------------
\subsection{~Implications for eROSITA}
\label{s5_eROSITA_implic}

\noindent
The next generation all-sky X-ray survey will be conducted by the upcoming eROSITA mission \cite{Predehl2010a}, which 
has an expected average survey PSF of 25\arcsec-30\arcsec and hence provides a lower spatial resolution than {\it Chandra} and XMM-{\it Newton}. In order to evaluate the performance for distant cluster applications, the high-$z$ AGN excess close to the cluster position is of critical importance. 

The  measured average excess of about one {\it full-band}-detected AGN within R$_{200}$ with a centrally peaked radial profile may pose a severe potential bias for X-ray spectroscopic applications  {at $z\!>\!0.9$} , most notably {high-$z$} ICM temperature measurements.  For faint and compact high-$z$ systems,  the full detected X-ray emission within the extraction aperture has to be considered to obtain enough signal, without the possibility to subtract embedded point sources.  Hidden AGN can hence harden the spectrum and bias the T$_X$ determination high \cite[see e.g.][]{Hilton2010a}.

%Since the measured  {\it soft-band} radial profile did not reveal a  significant central excess of AGN,  {\bf although the statistical uncertainties in the central region are still large}, the situation for the detection of extended high-$z$ Xray sources and accurate  {\it soft-band} flux determinations is still very promising. 

{Although the statistical uncertainties in the central region are still large, the measured  {\it soft-band} radial profile did not reveal a  significant central excess of AGN. The situation for the detection of extended high-$z$ Xray sources and accurate  {\it soft-band} flux determinations hence still looks  very promising. 
}
These tasks can be  performed in the soft-band only, most commonly in the 0.5-2\,keV range, where the cluster AGN contribution seems to be low for the presented X-ray selected sample {and sensitivities, which correspond to the planned eROSITA deep fields.}  
The detectability of distant clusters with eROSITA  {may} hence not be significantly  influenced by the present findings. The measured soft-band X-ray luminosity  for high-$z$ eROSITA clusters will hence likely be the preferable low-bias  mass proxy.

%applications:
%eROSITA
%
%a) resolution
%
%
%b) effect of centrally peaked low luminosity AGN
%
%c) on spectroscopy
%
%d) detection
%
%but: X-ray starting

%%%%%%%%%%%%%%%%%%%%%%%%%%%%%%%%%%%%%%%%%%%%%%%%%%%%%%%%
\section{~Summary and Conclusions}
\label{s6_Summary}
\noindent
This work investigated the X-ray point source excess in the environment of distant X-ray luminous galaxy clusters in the redshift range 
$0.9\!<\!z\!\la\!1.6$ based on a sample of 22 systems with an average mass of $2\times\!10^{14}\,\mathrm{M_{\sun}}$ compiled by the XMM-{\it Newton} Distant Cluster Project. The X-ray source detection in the available XMM-{\it Newton} observations for each cluster  with a median clean field exposure time of 19\,ksec were performed in two energy ranges, a 0.35-2.4\,keV {\it soft-band} and a 0.3-7.5\,keV {\it full-band}. The X-ray source counts were stacked in cluster-centric coordinates and compared to the average background counts extracted from three independent control fields in the XMM-{\it Newton}  field-of-view.
The main findings of this X-ray stacking analysis can be summarized as follows:

\begin{enumerate}
\item The cumulative radial X-ray source counts within a cluster-centric region of 8\arcmin \ reveal a significant excess of $\sim$67 ({\it soft-band}) and  $\sim$78 ({\it soft-band}) sources with a statistical confidence of 4.2\,$\sigma$ in the   {\it soft-band} and  4.0\,$\sigma$  in the {\it full-band}. The resulting average detected point source excess of 3.0$\pm$0.7 ({\it soft-band})  and  3.5$\pm$0.9 ({\it full-band}) sources per cluster environment are to be interpreted as lower limit due an incomplete coverage of the full geometric area out to  8\arcmin.

\item The radial cumulative distributions of the detected excess counts show a different radial behavior for the  {\it soft-band}  and {\it full-band}.  {\it Soft-band}  detected excess sources are mainly located at cluster-centric projected distances of 2-3\,Mpc (the 4\arcmin-6\arcmin-hump), while the {\it full-band} counts show an additional steep rise at small distances of r$<$2\arcmin.

\item The radial profiles of the background subtracted excess counts confirm the statistical significance of both radial features. The centrally peaked point source excess within the average cluster radius R$_{200}$ is consistent with various previous studies, most of them conducted at lower redshifts. The detected (incomplete) average excess per cluster is found to be about one point source per system with evidence for  lower X-ray luminosities below the  {\it soft-band}  detection limit. 

\item A second analysis was performed with all source lists rotated about the cluster center in a way that the principal elongation axes of the cluster emission are aligned for all systems. With the interpretation that the cluster elongation points in the direction of the main cluster assembly axis the tentative conclusion can be derived that the observed AGN  activity is mostly occurring along these identified matter infall directions.  

\item The second outer radial feature at cluster-centric distances  of 2-3\,Mpc was previously reported also for an intermediate-redshift cluster sample by \citet{Ruderman2005a}. However, a systematic boosting of this feature by foreground structures cannot be fully ruled out with the statistics of the present sample. The cross-correlation of public spectroscopic redshift information with the X-ray data, on the other hand, confirms the presence of a significant population of X-ray AGN in the cluster environments beyond a projected distance of 2\,Mpc. The observed 4\arcmin-6\arcmin-hump feature is consistent  with a population of bright     
{\it soft-band} -detected AGN triggered at a project distance of about 3$\times$R$_{200}$.

\item Taking all results at face value lends support to the idea of two different AGN populations and triggering mechanisms of nuclear activity in the distant cluster environments. In this picture, the fiducial cluster regions inside R$_{200}$ of high-$z$ X-ray luminous systems harbor low-luminosity AGN triggered by close galaxy encounters in the infall regions, while the excess in the outer cluster environment at distances of 2-3\,Mpc is due to major merger induced AGN activity.

\item With respect to distant cluster applications with the upcoming all-sky survey eROSITA, the results suggest that the spectroscopic temperature analysis of samples of distant cluster sources may be biased high due to embedded point sources, while the detection and flux measurements in the soft-band %should 
may not be significantly influenced in the general case.

\end{enumerate}

\noindent
The presented results are based on the currently largest homogeneously selected sample of X-ray luminous clusters at $z\!>\!0.9$. Statistical improvements and tests of the discussed results will require significantly larger, well-defined samples of distant X-ray clusters  of about 50 objects, which should  soon be available. The ultimate experiment of X-ray AGN in distant cluster environments at r$>$1\,Mpc will be made possible by eROSITA in the near future.

%implications
%outlook for larger samples

\acknowledgments 

\noindent
We thank the anonymous referee  for insightful comments that helped to improve the clarity of the paper.
We thank the XDCP team members for carrying out the distant cluster survey. 
RF would like to thank Gabriel Pratt for his comments, Andrea Merloni for helpful discussions, and Angela Bongiorno for the invitation to the special issue.
This research was supported by the DFG cluster of excellence `Origin and Structure of the Universe' (www.universe-cluster.de)
and  by the DFG under grants  BO 702/16-3. %, and the German DLR under grant 50 QR 0802. 
The XMM-{\it Newton} project is an ESA Science Mission with instruments and contributions directly funded by ESA Member
States and the USA (NASA). 
This research has made use of the NASA/IPAC Extragalactic Database (NED) which is operated by the Jet Propulsion Laboratory, California Institute of Technology, under contract with the National Aeronautics and Space Administration.

%\bibliography{bibliography}
\bibliographystyle{aa} % style aa.bst with Journal Definitions
\bibliography{../BIB/RF_BIB_11}

\end{document}